\documentclass[tighten,floats,aps,preprintnumbers]{revtex4}
\usepackage{amsmath}
\usepackage{epsf}
\usepackage{color}

\usepackage{ulem}

\newcounter{nombre}

\newcommand{\kb}{q}
\newcommand{\Chi}{{\cal X}}

\begin{document}
\preprint{KUNS-2333, RIKEN-MP-19}
\title{Alpha-cluster structure and density wave in oblate nuclei}

\author{Yoshiko Kanada-En'yo}
\affiliation{Department of Physics, Kyoto University,
Kyoto 606-8502, Japan}

\author{Yoshimasa Hidaka}
\affiliation{Mathematical Physics Lab., RIKEN Nishina Center, Saitama 351-0198, Japan}

\begin{abstract}
Pentagon and triangle shapes in $^{28}$Si and $^{12}$C are discussed in relation with 
nuclear density wave.
In the antisymmetrized molecular dynamics calculations,
the $K^\pi=5^-$ band in $^{28}$Si and the $K^\pi=3^-$ band in $^{12}$C are described
by the pentagon and triangle shapes, respectively.
 These negative-parity bands can be interpreted as the parity partners
of the  $K^\pi=0^+$ ground bands and they are constructed from the parity-asymmetric-intrinsic states.
The pentagon and the triangle shapes originate in $7\alpha$ and $3\alpha$ cluster structures,
respectively. 
In a mean-field picture, they are described also by the static 
one-dimensional density wave at the edge of the oblate states.
In analysis with ideal $\alpha$ cluster models using Brink-Bloch cluster wave functions
and that with a simplified model, 
we show that the static edge density wave for the pentagon and triangle shapes can be 
understood by spontaneous breaking of axial symmetry, i.e., the instability of the oblate states 
with respect to the edge density wave.
The density wave is enhanced in the $Z=N$ nuclei due to the proton-neutron coherent 
density waves, while it is suppressed in $Z\ne N$ nuclei.
\end{abstract}

\maketitle

\noindent
\section{Introduction}
\begin{figure}[th]
\epsfxsize=7 cm
\centerline{\epsffile{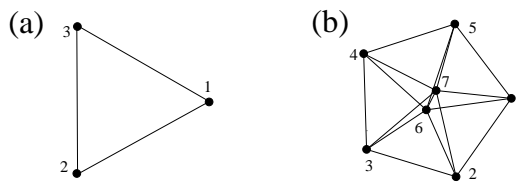}}
\caption{The schematic figures for spatial configurations of cluster centers of 
(a) a triangle structure consisting of three $\alpha$ clusters in $^{12}$C and (b) a pentagon structure of seven $\alpha$ clusters in$^{28}$Si.
\label{fig:penta}
}
\end{figure}

In light nuclei, some of negative-parity rotational bands with high $K$ quanta
are discussed in relation with specific symmetry of intrinsic states.
One of the famous examples is the $3^-$ state at $9.64$ MeV in $^{12}$C,  which has been discussed 
for a long time in connection to an equilateral triangle configuration of $3\alpha$ cluster structure. 
The $3^-$ state is the lowest negative-parity state, and its spin is contradict to the 
naive expectation from shell model calculations.
The reason for the low-lying $3^-$ state is understood by 
the point group $D_{3h}$ symmetry of the equilateral triangle $3\alpha$ 
cluster structure~\cite{brink66,yukawa70,horiuchi87} [see Fig.~\ref{fig:penta}(a)], 
which is characterized by the $n$-fold symmetry with $n=3$ 
of the intrinsic structure. The $3^-$ state is interpreted as the band head of the $K^\pi=3^-$ band 
constructed by the parity and total-angular-momentum projection from the 
$D_{3h}$ symmetry of the intrinsic state. 
Although the $4^-$ state in the $K^\pi=3^-$ band 
has not yet confirmed, a possible assignment $4^-$ for the level at $13.35$ MeV was suggested~\cite{nucldata}.
In microscopic $3\alpha$ cluster models, the $K^\pi=3^-$ band is considered to form a parity doublet 
with the $K^\pi=0^+$ ground  band~\cite{uegaki77}.

In $^{28}$Si, the $K^\pi=5^-$ rotational band starting from the $5^-_1$ state at $9.70$ MeV
was reported in $\gamma$-ray measurements by Glatz et al.~\cite{glatz81}, 
and it was discussed with a $7\alpha$ cluster structure with a pentagon shape~\cite{bauhoff82,bauhoff82b}. 
In the $7\alpha$ cluster model with Brink-Bloch (BB) $\alpha$-cluster wave functions~\cite{brink66}, an oblate solution for negative parity
shows the $D_{5h}$ symmetry [see Fig.~\ref{fig:penta}(b)]. Since a $K^\pi=5^-$ band can be constructed from the $D_{5h}$ symmetry of the intrinsic state, 
the existing $K^\pi=5^-$ band might be an indirect evidence of the pentagon shape and may be regarded as the 
parity partner of the $K^\pi=0^+$ ground state. 

In spite of the reasonable description of the $K^\pi=5^-$ band in $^{28}$Si 
with the $7\alpha$ cluster model,
the BB $\alpha$-cluster model is too simple to quantitatively describe 
low-lying energy spectra of $^{28}$Si~\cite{bauhoff82b}. Moreover, 
the validity of the ansatz that $^{28}$Si consists of seven $\alpha$ clusters is not obvious but it should be
checked with frameworks without any cluster assumptions 
because $\alpha$ clusters might be dissociated or melted down due to the spin-orbit force in $sd$-shell 
nuclei~\cite{Maruhn:2006ig}.

Recently, more sophisticated calculations of $^{28}$Si were performed  by one of the authors with 
antisymmetrized molecular dynamics (AMD)~\cite{KanadaEn'yo:2004cv,KanadaEn'yo:2003di}, 
which is a framework free from cluster assumptions.
The calculations reproduce well low-lying positive-parity levels of the oblate ground band 
and the excited prolate band by incorporating the spin-orbit force with a proper strength.
Interestingly, the $K^\pi=5^-$ band is constructed from the oblate state with a pentagon shape
even though existence of any clusters is not assumed in the 
calculations~\cite{KanadaEn'yo:2003di}.
This means that the pentagon shape can be induced by $\alpha$ cluster correlation
in the oblate intrinsic state of $^{28}$Si.

From the viewpoint of the symmetry breaking, the pentagon shape in the intrinsic state is 
regarded as the spontaneous breaking of the axial symmetry.
The surface density is oscillating along the edge of the oblate shape, that is,
the spontaneous symmetry breaking (SSB) occurs in the rotational invariance around
the symmetric axis. 
In relation to the SSB concerning density,   
this structure is associated with density wave (DW) in a nuclear matter,
in which the SSB of the translational invariance occurs.  
The DW in nuclear matter has been discussed for a long time~\cite{overhauser60,llano79,ui81},
and it was suggested that the one-dimensional DW could be stable 
in a low density nuclear matter~\cite{ui81}.
The nonuniform nuclear matter with density oscillation has also been investigated
with cluster models as the $\alpha$-cluster matter~\cite{brink73,tohsaki89,takemoto04}.
In analogy to the nuclear matter DW, 
the pentagon shape can be interpreted as the static one-dimensional DW
at the edge of the oblate state.
Our aim is to give description of the pentagon and triangle shapes in terms of the DW
with the wave number five and three to discuss the relation of the $7\alpha$ and $3\alpha$ 
cluster structures with the static edge DW, i.e., the spontaneous axial symmetry breaking 
of the oblate states.

In this paper, we report the AMD calculations of $^{28}$Si while focusing on the pentagon shape
in the oblate states. In analysis of single-particle wave
functions of the obtained AMD wave functions, we show that 
the pentagon shape is expressed by 
one-particle and one-hole excitations having the wave number five on the oblate state
and it can be interpreted as the one-dimensional DW.
To see the instability of the axial symmetry
with respect to the pentagon shape,
analyses of ideal cluster models using BB wave functions
are performed. 
Similarly, focusing on the triangle shape of the $3\alpha$ cluster structure, 
 structures of $^{12}$C are also discussed.
We introduce a simplified model for the one-dimensional DW in the oblate state
by truncating active orbits for particle and hole states, 
and show that the proton-neutron coherent DWs in $Z=N$ nuclei
promote the instability of the oblate states with respect to the pentagon and triangle shapes.
The suppression mechanism of cluster structures in neutron-rich nuclei 
is discussed from the viewpoint of proton DW with excess neutrons.

The paper is organized as follows: 
In the next section, we explain the AMD calculations for $^{28}$Si and $^{12}$C. 
Analyses with ideal cluster models using BB $\alpha$-cluster wave functions
are given in Sec.~\ref{sec:analysis}, and those using extended BB wave functions in Sec.~\ref{sec:analysis2}.
Discussions with a simplified model for the one-dimensional DW
are given in Sec.~\ref{sec:discussion}.
Finally, in Sec.~\ref{sec:summary}, a summary and an outlook are given.

\section{AMD calculations of $^{28}$Si and $^{12}$C}
\subsection{Method of AMD calculations}\label{subsec:AMD}
The AMD method has been applied for various nuclei 
and succeeded to describe shell-model structures and cluster structures 
of ground and excited states in a light-mass region~\cite{ENYOabc,ENYOsupp,AMDrev}.
Here we briefly describe a simple version of the AMD method and its application to
$^{28}$Si and $^{12}$C~\cite{ENYOabc,KanadaEn'yo:2004cv} focusing on cluster features of
the oblate bands.
The details of the previous AMD calculations for $^{28}$Si are described in 
Refs.~\cite{KanadaEn'yo:2004cv,KanadaEn'yo:2003di}, in which 
the energy levels of the $K^\pi=0^+_1$, $K^\pi=0^+_2$, 
$K^\pi=3^-$ and, $K^\pi=5^-$ bands are well reproduced.

An AMD wave function for an $A$-nucleon system 
is given by a Slater determinant of 
Gaussian wave packets,
\begin{equation}
 \Phi_{\rm AMD}({\bf Z}) = \frac{1}{\sqrt{A!}} {\cal{A}} \{
  \varphi_1,\varphi_2,...,\varphi_A \},
\end{equation}
where the $i$-th single-particle wave function is written as
\begin{align}
 \varphi_i&= \phi_{{\bf Z}_i}\Chi_i,\label{eq:single}\\
 \phi_{{\bf Z}_i}({\bf r}_j) &\propto
\exp\Bigl[-\nu\Bigl({\bf r}_j-\frac{{\bf Z}_i}{\sqrt{\nu}}\Bigr)^2\Bigr].
\label{eq:spatial}
\end{align}
$\Chi_i$ is the spin-isospin function and 
fixed to be $p\uparrow,$ $p\downarrow$, $n\uparrow$, or $n\downarrow$. 
The spatial part is represented by 
complex variational parameters, ${\rm Z}_{xi}$, ${\rm Z}_{yi}$, 
${\rm Z}_{zi}$, which indicate the centers of the $i$-th Gaussian wave packet. 
The parameter $\nu$ is chosen to be 
$\nu=0.15$ fm$^{-2}$ and $\nu=0.175$ fm$^{-2}$ for $^{28}$Si and
$^{12}$C so as to minimize the energy of the positive-parity state. 

In the AMD model, all the centers $\{{\bf Z}_1,{\bf Z}_2,\cdots,{\bf Z}_A\}$ of single-nucleon Gaussians are treated independently as complex variational parameters. Thus, the AMD method is based completely on single
nucleons and therefore it is free from such assumptions 
as cluster existence or axial symmetry.
Nevertheless, if a cluster structure is favored in a system, 
the cluster structure can be described as an optimum solution of
AMD wave functions because BB cluster wave functions 
are included in the AMD model space. For instance, $\alpha$ cluster formation
is expressed by the concentration of Gaussian centers for four nucleons,
$p\uparrow$, $p\downarrow$, $n\uparrow$, and $n\downarrow$ at a certain position.

By using an effective Hamiltonian, 
\begin{equation}
H_{\rm eff}=\sum_i T_i +\sum_{i<j} v_{ij} +\sum_{i<j<k} v_{ijk}
\end{equation}
consisting of kinetic terms and two-body and three-body interaction terms as effective nuclear forces, 
the energy variation is performed within the AMD model space to 
obtain the optimum solutions, which correspond to the intrinsic wave functions for low-lying states.
As in Refs.~\cite{ENYOabc,KanadaEn'yo:2004cv,KanadaEn'yo:2003di}, 
the energy variation is done after parity projection by operating $(1\pm P_r)$ on
the AMD wave function. 
After the energy variation for $(1\pm P_r)\Phi_{\rm AMD}({\bf Z})$
with respect to ${\bf Z}$ the optimized intrinsic wave functions, $\Phi_{\rm AMD}({\bf Z}^{(+)})$  and 
$\Phi_{\rm AMD}({\bf Z}^{(-)})$, are obtained for the positive- and negative-parity 
states, respectively. Then, the total-angular-momentum projection, $P^J_{MK}$, is operated on 
the obtained AMD wave functions, $P^J_{MK} (1\pm P_r)\Phi_{\rm AMD}$, to calculate
expectation values of parity and angular-momentum  eigenstates, $J^\pm$.

The adopted effective nuclear forces are the same as those in 
Refs.~\cite{KanadaEn'yo:2004cv,KanadaEn'yo:2003di} with which AMD 
calculations reproduce the energy levels of $^{28}$Si. 
Namely, the MV1 force (case 1)~\cite{TOHSAKI}, which consists of finite-range two-body 
and zero-range three-body forces,
 with a parameter set $(b=h=0, m=0.62)$ is used for the central force. As for the spin-orbit force, 
the spin-orbit term of the G3RS force~\cite{LS} with the strengths $u_{I}=-u_{II}=2800$ MeV is adopted.
Coulomb force is approximated by
seven-range Gaussians. 

\subsection{AMD results of $^{28}$Si}
\begin{figure}[th]
\epsfxsize=5.5 cm
\centerline{\epsffile{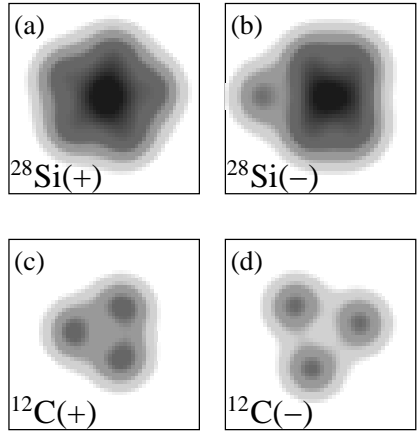}}
\caption{Density distributions of the AMD wave functions for the positive- and negative-parity
states in $^{28}$Si and$^{12}$C. 
\label{fig:dense}
}
\end{figure}

\begin{figure}[th]
\epsfxsize=7 cm
\centerline{\epsffile{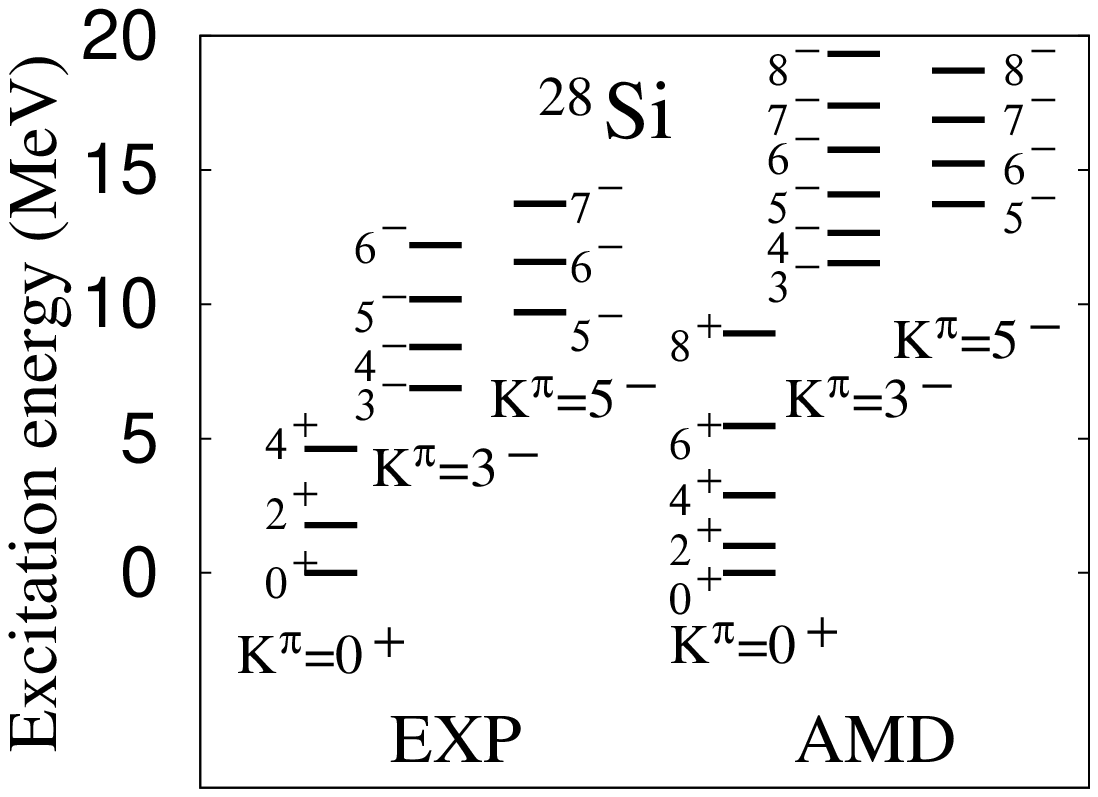}}
\epsfxsize=7 cm
\centerline{\epsffile{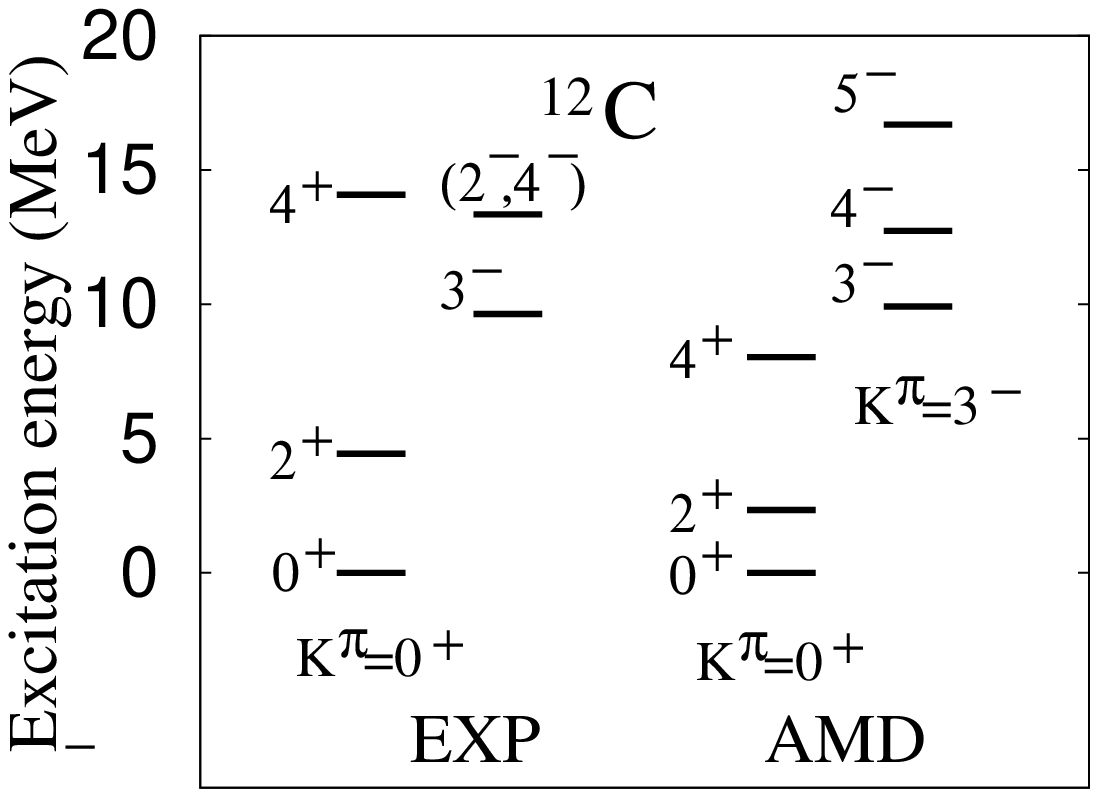}}
\caption{(a) Energy levels calculated with AMD calculations and the experimental levels of the
$K^\pi=0^+_1$,  $K^\pi=3^-_1$, and $K^\pi=5^-_1$ bands in $^{28}$Si. (b) Those 
of the $K^\pi=0^+_1$ and  $K^\pi=3^-_1$ bands  in  $^{12}$C. The experimental data are taken from 
Refs.~\cite{nucldata,glatz81}.
\label{fig:spe}
}
\end{figure}

In the present work, we focus only on the oblate rotational bands, $K^\pi=0^+_1$,  $K^\pi=3^-_1$, and $K^\pi=5^-_1$,
though the prolate excited $K^\pi=0^+_2$ band exists in $^{28}$Si~\cite{KanadaEn'yo:2004cv}.
We adopt the oblate solutions of the intrinsic wave functions $\Phi_{\rm AMD}({\bf Z}^{(+)})$ and $\Phi_{\rm AMD}({\bf Z}^{(-)})$ 
obtained by the energy variation after positive- and negative-parity projections. 
$\Phi_{\rm AMD}({\bf Z}^{(+)})$ and $\Phi_{\rm AMD}({\bf Z}^{(-)})$ correspond to the intrinsic states of the
lowest positive- and negative-parity bands.  
Density distributions of these AMD wave functions are shown in Fig.~\ref{fig:dense}.
Interestingly, the wave function $\Phi_{\rm AMD}({\bf Z}^{(+)})$, which corresponds to the $K^\pi=0^+$ ground band,
shows the pentagon shape due to the $7\alpha$-like structure even though $\alpha$ clusters are not {\it a priori} 
assumed in the framework. We should comment that $\alpha$ clusters in $\Phi_{\rm AMD}({\bf Z}^{(+)})$
are somehow dissociated due to the spin-orbit force as discussed in Ref.~\cite{KanadaEn'yo:2004cv}. 
From these two intrinsic states $\Phi_{\rm AMD}({\bf Z}^{(+)})$ and 
$\Phi_{\rm AMD}({\bf Z}^{(-)})$, we calculate
the $J^\pm$ states by performing the parity and angular-momentum projections and diagonalizing Hamiltonian and norm matrices with respect to $P^J_{MK} (1\pm P_r)\Phi_{\rm AMD}({\bf Z}^{(+)})$ and $P^J_{MK} (1\pm P_r)\Phi_{\rm AMD}({\bf Z}^{(-)})$.
Here, states with each parity are described by a linear combination of 
the parity and angular-momentum eigenstates projected from both of $\Phi_{\rm AMD}({\bf Z}^{(+)})$ and $\Phi_{\rm AMD}({\bf Z}^{(-)})$. 
The calculated energy levels of $^{28}$Si are shown in Fig.~\ref{fig:spe} 
compared with the experimental data of the members in the oblate bands, $K^\pi=0^+_1$, $K^\pi=3^-_1$, and $K^\pi=5^-_1$.
The experimental energy levels are reproduced rather well by the calculations.
The calculated in-band $E2$ transition strengths also reproduce well the experimental data (see Table.~\ref{tab:be2}).  
The ground band, $K^\pi=0^+_1$, and the lowest negative-parity band, $K^\pi=3^-_1$, are dominantly constructed 
from $\Phi_{\rm AMD}({\bf Z}^{(+)})$ and $\Phi_{\rm AMD}({\bf Z}^{(-)})$, respectively. 
The $K^\pi=5^-_1$ band is mainly constructed from $\Phi_{\rm AMD}({\bf Z}^{(+)})$ having the pentagon shape
though the $K^\pi=5^-_1$ band member states have significant mixing with the $K^\pi=3^-_1$ band members. 
This means that the $K^\pi=0^+_1$ and the $K^\pi=5^-_1$ bands can be interpreted as the parity partners
constructed from the parity-asymmetric-intrinsic state $\Phi_{\rm AMD}({\bf Z}^{(+)})$ with the pentagon shape.

\begin{table}[ht]
\caption{\label{tab:be2} The calculated and the experimental values of 
$E2$ transition strengths in $^{28}$Si.
The values in Weisskopf unit,
W.u.=5.05 e$^2$fm$^{-4}$ are listed.
The $B(E2)$ values are calculated with the AMD method. 
The experimental data are taken from Ref.~\cite{glatz81}.}
\begin{center}
\begin{tabular}{cccc}
\hline
{initial} & {final} & \multicolumn{2}{c}{$B(E2)$}\\
$J^\pm_f$ & $J^\pm_i$ &exp.& cal. \\ 
\hline
\multicolumn{2}{c}{$K^\pi=0^+_1\rightarrow 0^+_1$} & & \\
$2^+_1$	&	$0+_1$	&$	12.7	(+0.4,-0.3)	$&	10.6 	\\
$4^+_1$	&	$2^+_1$	&$	13.6	(+1.4,-1.2)	$&	15.1 	\\
$6^+_1$	&	$4^+_1$	&$	9.4	(+3.6,-2.0)	$&	16.2 	\\
\hline
\multicolumn{2}{c}{$K^\pi=3^-_1\rightarrow  3^-_1$} & & \\
$4^-_1$	&	$3^-_1$	&$	32.4	(+9.6,-6.4)	$&	22.4 	\\
$5^-_2$	&	$3^-_1$	&$	>3.4		$&	2.0 	\\
$5^-_2$	&	$4^-_1$	&$			$&	13.7 	\\
$6^-_2$	&	$4^-_1$	&$	>6.1		$&	2.5 	\\
$6^-_2$	&	$5^-_2$	&$	>12		$&	16.9 	\\
$7^-_2$	&	$5^-_2$	&$			$&	5.2 	\\
$7^-_2$	&	$6^-_2$	&$			$&	16.9 	\\
$8^-_2$	&	$6^-_2$	&$			$&	7.8 	\\
\hline
\multicolumn{2}{c}{$K^\pi=5^-_1\rightarrow 5^-_1$} & & \\
$6^-_1$	&	$5^-_1$	&$	17	(+8,-4)	$&	14.4 	\\
$7^-_1$	&	$5^-_1$	&$	>2.5		$&	7.7 	\\
$7^-_1$	&	$6^-_1$	&$	>16.5		$&	13.8 	\\
$8^-_1$	&	$6^-_1$	&$			$&	10.1 	\\
\hline
\multicolumn{2}{c}{$K^\pi=5^-_1\rightarrow 3^-_1$} & & \\
$5^-_1$	&	$3^-_1$	&$	0.034	(+0.01,-0.01)	$&	4.0 	\\
$5^-_1$	&	$4^-_1$	&$	2	(+0.2,-0.2)	$&	7.1 	\\
$7^-_1$	&	$5^-_2$	&$	>2.3		$&	2.8 	\\
$6^-_1$	&	$4^-_1$	&$			$&	7.7 	\\
\hline
\end{tabular}
\end{center}
\end{table}

Next we analyze the single-particle wave functions in the pentagon intrinsic state. 
The single-particle wave functions $\varphi_i$ in Eq.~(\ref{eq:single}) written by Gaussian wave packets are non-orthogonal
to each other. We can make linear transformation from the set $\{\varphi_i\}$ to an orthonormal basis set $\{\varphi'_i\}$ keeping the Slater determinant unchanged 
except for normalization, ${\rm det}\{\varphi'_i\}\propto{\rm det}\{\varphi_i\}=\Phi_{\rm AMD}({\bf Z})$.
From this orthonormal basis, we construct the Hartree-Fock(HF) single-particles $\{\varphi^\text{HF}
_i\}$, which diagonalize
the HF single-particle Hamiltonian as described in Ref.~\cite{DOTE-be}.
Analysis of  $\{\varphi^\text{HF}_i\}$ is helpful to discuss intrinsic states in a mean-field picture.

Among the HF single-particle wave functions in the intrinsic wave function, $\Phi_{\rm AMD}({\bf Z}^{(+)})$, 
we find single-particle orbits with pentagon density distributions (see Fig.~\ref{fig:single}).
The pentagon orbits show the parity asymmetry indicating mixing of positive-parity and negative-parity components.
We extract each parity component from the pentagon orbits, and find that 
about 5\% negative-parity component is mixed in the dominant positive-parity component in each orbit.
As shown in Fig.~\ref{fig:single}, both the positive- and negative-parity components 
show donut shapes in their density distributions, and the pentagon orbits
can be roughly described by a linear combination 
$c\phi_{(0,0,\pm 2)} +c'\phi_{(0,0,\mp 3)}$ with $|c'|^2\sim 0.05$ for $|c|^2+|c'|^2=1$, in terms
of harmonic oscillator (H.O.) single-particle orbits labeled by quantum numbers ($n_z$, $n_\rho$, $m_l$) 
in the cylinder coordinates (see Appendix \ref{appendix:ho} for the expressions of $\phi_{(0,0,\pm 2)}$ and $\phi_{(0,0,\pm 3)}$). 
In the pentagon state of $^{28}$Si, totally eight pentagon orbits are found 
corresponding to $c\phi_{(0,0,\pm 2)} +c'\phi_{(0,0,\mp 3)}$ occupied by four
species of nucleons, $p\uparrow$, $p\downarrow$, $n\uparrow$,
and $n\downarrow$.

\subsection{AMD results of $^{12}$C}
The calculations of $^{12}$C with the AMD are described in 
Ref.~\cite{ENYOsupp}, where the $K^\pi=0^+_1$ and $K^\pi=3^-_1$ bands are constructed from
triangle states with oblate deformations. 
In the present work, we use the same effective interactions as those for $^{28}$Si.

Both of the intrinsic wave functions $\Phi_{\rm AMD}({\bf Z}^{(+)})$ and $\Phi_{\rm AMD}({\bf Z}^{(-)})$
for positive and negative parity show equilateral triangle shapes because of the $3\alpha$ structure.
As seen in density distributions shown in Figs.~\ref{fig:dense}(c) and \ref{fig:dense}(d), the
development of the $3\alpha$-cluster structure is more remarkable in  
the negative-parity intrinsic state, $\Phi_{\rm AMD}({\bf Z}^{(-)})$, than the
positive-parity intrinsic state, $\Phi_{\rm AMD}({\bf Z}^{(+)})$. 
To calculate energy levels, we perform parity and angular-momentum projections and obtain 
the rotational $K^\pi=0^+_1$ and $K^\pi=3^-_1$ bands constructed from 
the triangle intrinsic states, $\Phi_{\rm AMD}({\bf Z}^{(+)})$ and $\Phi_{\rm AMD}({\bf Z}^{(-)})$,
respectively.
If we tolerate the difference of the degree of the cluster development between 
positive- and negative-parity states, 
the $K^\pi=0^+_1$ and $K^\pi=3^-_1$ bands are roughly interpreted as 
the parity partners of the parity-asymmetric-intrinsic state with the triangle shape
as argued in Ref.~\cite{uegaki77}.
Then, we can say that the triangle shape of $^{12}$C has an analogy to the pentagon
shape of $^{28}$Si constructing the parity partner $K^\pi=0^+_1$ and $K^\pi=5^-_1$ bands.

\begin{figure}[th]
\epsfxsize=7 cm
\centerline{\epsffile{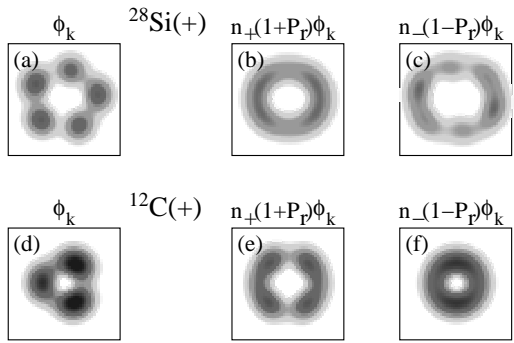}}
\caption{(a) Density of the highest single-particle orbit $\varphi^\text{HF}_k$ 
in $\Phi_{\rm AMD}({\bf Z}^{(+)})$ of  
$^{28}$Si. (b) and (c) Density for the positive- and negative-parity components of the highest orbit.
Each component is normalized to be 1 by multiplying $n_pm$ with $1/n^2_\pm= 
{\langle\varphi^\text{HF}_k|(1\pm P_r)^2|\varphi^\text{HF}_k\rangle}$.
(d) Density of the highest single-particle orbit in $\Phi_{\rm AMD}({\bf Z}^{(+)})$ for 
$^{12}$C. (e) and (f) Density for the positive and negative-parity components of the highest orbit.
Each component is normalized to be 1.
\label{fig:single}
}
\end{figure}

In a similar way to $^{28}$Si, we analyze the HF single-particles $\{\varphi^\text{HF}
_i\}$ in the $\Phi_{\rm AMD}({\bf Z}^{(+)})$ and find the parity-mixing single-particle orbits
with the triangle shape. The density distribution of a
$\varphi^\text{HF}_i$ for the triangle shape is shown in Fig.~\ref{fig:single}(d), and  
those of the positive- and negative-parity components extracted from 
this triangle orbit are shown in Figs.~\ref{fig:single}(e) and \ref{fig:single}(f). 
Both the positive- and negative-parity components 
show donut shape densities, and the triangle orbit can be roughly interpreted as a 
linear combination 
$c\phi_{(0,0,\pm 1)} +c'\phi_{(0,0,\mp 2)}$ with $|c'|^2\sim 0.06$ for $|c|^2+|c'|^2=1$, in
terms of H.O. single-particle orbits expressed by cylinder coordinates (see Appendix \ref{appendix:ho}).
In the triangle structure of $^{12}$C, totally eight triangle orbits are found 
corresponding to $c\phi_{(0,0,\pm 1)} +c'\phi_{(0,0,\mp 2)}$ occupied by 
$p\uparrow$, $p\downarrow$, $n\uparrow$, and $n\downarrow$.

\section{Analysis with Brink-Bloch $\alpha$ cluster models}\label{sec:analysis}

As described in the previous section, the pentagon and triangle shapes are found in the AMD results 
of $^{28}$Si and $^{12}$C, respectively. The AMD calculations show that 
these shapes originate in the $7\alpha$- and $3\alpha$-cluster features 
in which the $\alpha$ clusters are somehow dissociated because of the spin-orbit force. 
The fact that the $7\alpha$- and $3\alpha$-like structures 
are actually formed in the AMD without {\it a priori} assuming any clusters
indicates that these cluster structures are favored in 
$^{28}$Si and $^{12}$C.

The pentagon and triangle shapes are interpreted as spontaneous breaking of axial symmetry 
of the oblate $^{28}$Si and $^{12}$C.
In this section, to understand 
the mechanism of the symmetry breaking, 
we investigate properties of ideal $7\alpha$- and $3\alpha$-cluster states
by using BB $\alpha$-cluster wave functions~\cite{brink66}.

\subsection{Brink-Bloch $\alpha$-cluster wave functions}
BB $\alpha$-cluster wave functions $\Phi^{\rm BB}_{X\alpha}$ 
for even-even $Z=N$ nuclei with the mass number $A=4X$ are
described by $X\alpha$-cluster wave functions consisting of $(0s)^4$ $\alpha$ clusters~\cite{brink66,horiuchi87}.
The $i$-th $\alpha$ cluster is located around a certain position ${\bf S}_i$, and 
$\Phi^{\rm BB}_{X\alpha}$ 
is characterized by a spatial configuration of center positions of $X\alpha$ clusters,
$\{ {\bf S}_1,\cdots,  {\bf S}_X\}$.

A BB $\alpha$-cluster wave function  $\Phi^{\rm BB}_{X\alpha}$ for a $X\alpha$ state can be expressed also by an
AMD wave function with a specific configuration of Gaussian centers $\{{\bf Z}\}$. When 
$\Chi_i$ is chosen to be $p\uparrow$, $p\downarrow$, $n\uparrow$, and $n\downarrow$
for $i=\{1\cdots X\}$, $i=\{X+1,\cdots, 2X\}$, $i=\{2X+1,\cdots, 3X\}$, and $i=\{3X+1,\cdots, 4X\}$, respectively, and 
Gaussian centers for four 
nucleons ($p\uparrow$, $p\downarrow$, $n\uparrow$, $n\downarrow$) 
are common and real values,
${\bf Z}_i={\bf Z}_{i+X}={\bf Z}_{i+2X}={\bf Z}_{i+3X}={\bf S}_i/\sqrt{\nu}$ ($i=1 \cdots X$), 
the AMD wave function 
is equivalent to the corresponding BB $\alpha$-cluster wave function for 
$X$ $\alpha$ clusters localizing at the positions ${\bf S}_1, {\bf S}_2,\cdots,{\bf S}_X$. 

\subsection{Pentagon $7\alpha$ BB wave function}
Let us consider the pentagon structure of a $7\alpha$ system.  
The $\alpha$-cluster centers ${\bf S}_i$ of $\alpha$ clusters are taken to have the pentagon configuration 
illustrated in Fig.~\ref{fig:penta}(b) as 
\begin{equation}\label{eq:BB-penta1}
{\bf S}_i= \Bigl(d\sqrt{\nu} \cos\Bigl(\frac{2\pi}{5}i\Bigr), d\sqrt{\nu} \sin\Bigl(\frac{2\pi}{5}i\Bigr), 0\Bigr)
\end{equation} 
for $i=1,\cdots, 5$, and
\begin{equation}\label{eq:BB-penta2}
{\bf S}_i= (0,0,\pm d'\sqrt{\nu}) 
\end{equation} 
for $i=6,7$. $d$ is the dimensionless pentagon size. Thus, the defined $\Phi^{\rm BB}_{7\alpha}$ is determined by three parameters $\nu, d, d'$,
and hence, we denote the pentagon $7\alpha$ wave functions by $\Phi^{\rm BB}_{7\alpha}(\nu, d, d')$.

Next we explain the relation between $\Phi^{\rm BB}_{7\alpha}$ and shell model wave functions
by transforming the single-particle wave functions of $\Phi^{\rm BB}_{7\alpha}$ 
in the expansion with respect to the pentagon size $d$.
In general, when $\alpha$-cluster centers $\{ {\bf S}_i \}$ are located around the origin, 
the BB wave function can be connected to a H.O. shell model
wave function by using invariance of a Slater determinant
$\det\{\phi_i({\bf r}_j)\} =n_0\det\{\phi'_i({\bf r}_j)\}$
under a linear transformation $\phi_{i}({\bf r})\rightarrow \phi'_{i}({\bf r})$. Here
$n_0$ is a normalization factor.

For oblate systems, we use H.O. single-particle wave functions 
$\phi_{(n_z,n_\rho,m_l)}$ in the expression of cylinder coordinates described in Appendix \ref{appendix:ho}. 
In the small $d'$ limit, the spatial wave functions for spin-up protons in $\Phi^{\rm BB}_{7\alpha}(\nu, d, d')$ can be transformed to $\det\{\phi'_i({\bf r}_j)\}$, which is given
by the Taylor expansion with respect to the pentagon size $d$ as follows: 
\begin{align}\label{eq:7alpha-d-expansion}
\det\{\phi_{{\bf Z}_1}, \phi_{{\bf Z}_2}, \cdots, \phi_{{\bf Z}_7}\} &=
n_0\det\{\phi'_1, \phi'_2, \cdots, \phi'_7\}, \notag \\
 \phi'_1&=\phi_{(0,0,0)} + O(d^2), \notag \\
 \phi'_2&=\phi_{(1,0,0)} + O(d^2), \notag \\
 \phi'_3&=\phi_{(0,0,+1)} + O(d^2), \notag \\
 \phi'_4&=\phi_{(0,0,-1)} + O(d^2), \notag \\
 \phi'_5&=\phi_{(0,1,0)} + O(d^2), \notag \\
 \phi'_6&=\phi_{(0,0,-2)} - \frac{d}{\sqrt{6}} \phi_{(0,0,+3)} + O(d^2),  \notag \\
 \phi'_7&=\phi_{(0,0,+2)} + \frac{d}{\sqrt{6}} \phi_{(0,0,-3)} + O(d^2),
\end{align}
where $n_0$ has the order $O(d^9,d')$.
We define 
$\phi'^{(0)}_{1,2,3,4,5,6,7}\equiv\phi_{(0,0,0)}$, 
$\phi_{(1,0,0)}$, $\phi_{(0,0,+1)}$, $\phi_{(0,0,-1)}$, $\phi_{(0,1,0)}$, $\phi_{(0,0,+2)}$, $\phi_{(0,0,-2)}$.
In the small $d$ limit, single-particle orbits $\phi'_i$ approach $\phi'^{(0)}_i$ and 
the $7\alpha$ wave function becomes equivalent to the following $0\hbar\omega$ shell model wave function
with the $s_\pi^2 s_\nu^2 p_\pi ^6 p_\nu^6 (sd)_\pi^6 (sd)_\nu^6$ configuration,
\begin{equation}
\Phi^{\rm BB}_{7\alpha}(\nu,d',d) \rightarrow \Phi^{(0\hbar\omega)}_{7\alpha}\equiv
n_0^4\prod_{\tau\sigma} 
\det\{\phi'^{(0)}_1 \Chi_{\tau\sigma}, \cdots, \phi'^{(0)}_7\Chi_{\tau\sigma}
\},
\end{equation}
where $\tau=\{p,n\}$ and $\sigma=\{\uparrow,\downarrow\}$.
In this limit, $\Phi^{(0\hbar\omega)}_{7\alpha}$ has the axial symmetric oblate shape. 
Here after we consider only a small $d'$ limit and investigate properties of 
$\Phi^{\rm BB}_{7\alpha}(\nu, d, d')$ as functions of $\nu$ and $d$, $\Phi^{\rm BB}_{7\alpha}(\nu, d)$.
In particular, the symmetry breaking of the oblate shape caused by the finite $d$ is discussed. 

Le us consider the pentagon shape described by $\Phi^{\rm BB}_{7\alpha}$ with 
a finite pentagon size $d$. As the pentagon size $d$ increases, the pentagon shape develops and
the axial symmetry breaking of the oblate state enlarges. 
The leading terms of the deviation from 
$\Phi^{(0\hbar\omega)}_{7\alpha}$ are contained in 
$\phi'_6$ and $\phi'_7$.
The orbits $\phi'_6$ and $\phi'_7$ are the 
parity-mixed orbits, and they show density with the pentagon shape
as expressed in the following explicit form of the density,
\begin{align}\label{eq:cos5phi}
 \phi'^*_6({\bf r})   \phi'_6({\bf r}) &=\phi'^*_7({\bf r}) \phi'_7({\bf r})
\notag\\
&= \frac{1}{2(\pi b^2)^{3/2}} \left( \frac{\rho}{b}\right)^4
e ^{-r^2/b^2} \left( 1+\frac{d}{3\sqrt{2}}\frac{\rho}{b}\cos(5\phi) +O(d^2) \right).
\end{align}
The second term $\cos(5\phi)$ gives the density oscillation with the wave number five along the edge of the oblate shape and show the pentagon feature.
The orbits $\phi'_6$ and $\phi'_7$ are given by linear combinations of 
$\phi_{(0,0,\mp 2)}$ and $\phi_{(0,0,\pm 3)}$. Due to the mixing of $\phi_{(0,0,\pm 3)}$
in $\phi_{(0,0,\mp 2)}$ of the amplitude $d^2/6$,
the density of these orbits changes from the axial symmetric density to the oscillating density.
This is nothing but the symmetry breaking of the rotational invariance around
the $z$ axis. If we associate the rotational invariance with the translational invariance of 
a uniform matter, and
the $z$-component of the angular momentum $m_l$ with the momentum $k$, then we find 
a good correspondence of the $\phi'_6$ and $\phi'_7$ orbits with the single-particle
wave functions of the nuclear matter DW proposed by Overhauser~\cite{overhauser60}.
In other words, the pentagon shape can be interpreted as 
the static DW at the edge of the oblate state.
As shown below, $\Phi^{\rm BB}_{7\alpha}(\nu,d)$ is expressed by coherent 
particle-hole configurations from 
$\Phi^{(0\hbar\omega)}_{7\alpha}$.

We show the particle-hole representation of $\Phi^{\rm BB}_{7\alpha}(\nu, d)$ below. 
We assume that $\Phi^{(0\hbar\omega)}_{7\alpha}$ is the Hartree-Fock vacuum $|0\rangle_{\rm F}$,
and $\phi_{(0,0,\pm 3)} \Chi_{\tau\sigma}$ and $\phi_{(0,0,\pm 2)} \Chi_{\tau\sigma}$ 
are the levels above and below the Fermi level, respectively. 
We define the particle and hole operators as
\begin{align}
a^\dagger_{\pm k,\tau\sigma} &= c^\dagger_{\pm k,\tau\sigma}, \notag\\
b^\dagger_{\pm \kb,\tau\sigma} &= c_{\mp \kb,\tau-\sigma}, 
\end{align}
where the labels $k \equiv3$ and $\kb \equiv 2$ indicate $m_l$ for particles and holes.
When higher order terms, $O(d^2)$, in the single-particle wave functions $\phi'$ are ignored, 
$\Phi^{\rm BB}_{7\alpha}(\nu, d)$ can be approximated to be 
\begin{equation}
\Phi^{\rm BB}_{7\alpha}(\nu, d)\approx \prod_{\chi}\left(1+\frac{d}{\sqrt{6}} a^\dagger_{-k,\chi} b^\dagger_{-\kb,-\chi} \right) 
\left(1-\frac{d}{\sqrt{6}} a^\dagger_{+k,\chi} b^\dagger_{+\kb,-\chi} \right)  |0 \rangle_{\rm F},
\label{eq:7alpha-1p1h}
\end{equation}
$\chi=\tau\sigma$ and $-\chi=\tau-\sigma$.
As clearly seen, the product of the particle and hole operators, 
$a^\dagger_{\pm k,\chi} b^\dagger_{\pm \kb,-\chi}$,
brings quanta $K= \pm 5$.

\subsection{Triangle $3\alpha$ BB wave function}
In a similar way to the pentagon $7\alpha$ state $\Phi^{\rm BB}_{7\alpha}(\nu,d)$,  
the equilateral triangle $3\alpha$ state is related to 
axial symmetry breaking of the oblate state in $p$-shell, and triangle shape is described by
parity-mixed orbits.
In the BB $\alpha$-cluster wave function $\Phi^{\rm BB}_{3\alpha}$ for the $3\alpha$ structure, 
the parameters ${\bf S}_i$ ($i=1,\cdots, 3$) with a triangle configuration are written as
\begin{equation}\label{eq:BB-triangle1}
{\bf S}_i= \Bigl(d\sqrt{\nu} \cos\Bigl(\frac{2\pi}{3}i\Bigr), d\sqrt{\nu} \sin\Bigl(\frac{2\pi}{3}i\Bigr), 0\Bigr).
\end{equation} 
$\Phi^{\rm BB}_{3\alpha}$ is specified by the parameters $\nu, d$ as
$\Phi^{\rm BB}_{3\alpha}(\nu, d)$. $d$ is the dimensionless triangle size.
With a proper transformation $\phi_{{\bf Z}_i}({\bf r})\rightarrow \phi'_{i}({\bf r})$ and
the Taylor expansion of the transformed single-particle orbits $\phi'_{i}({\bf r})$
with respect to the triangle size $d$, we can rewrite the spatial wave function for three identical nucleons,  
\begin{equation}\label{eq:3alpha-d-expansion}
\det\{\phi_{{\bf Z}_1}, \phi_{{\bf Z}_2}, \phi_{{\bf Z}_3}\} =
n_0\det\{\phi'_1, \phi'_2, \phi'_3\},
\end{equation}
with 
\begin{align}
 \phi'_1&=\phi'^{(0)}_1, \notag\\
 \phi'_2&=\phi'^{(0)}_2 + \frac{d}{2} \phi_{(0,0,+2)} + O(d^2), \notag\\
 \phi'_3&=\phi'^{(0)}_3 - \frac{d}{2} \phi_{(0,0,-2)} + O(d^2),
\end{align}
where 
\begin{align}
 \phi'^{(0)}_1&\equiv\phi_{(0,0,0)} ,\notag\\
 \phi'^{(0)}_2&\equiv\phi_{(0,0,-1)}, \notag\\
 \phi'^{(0)}_3&\equiv\phi_{(0,0,+1)}.
\end{align}
In the small $d$ limit, $\Phi^{\rm BB}_{3\alpha}(\nu,d)$ becomes equivalent to the
$0\hbar\omega$ shell model wave function,  
\begin{align}
\Phi^{\rm BB}_{3\alpha}(\nu, d) &\rightarrow n^4_0 \Phi^{(0\hbar\omega)}_{3\alpha}, \notag\\
\Phi^{(0\hbar\omega)}_{3\alpha} &\equiv 
\prod_{\tau \sigma} 
\det\{\phi'^{(0)}_1 \Chi_{\tau\sigma}, \cdots, \phi'^{(0)}_3\Chi_{\tau\sigma}
\}. 
\end{align}
The orbits $\phi'_2$ and $\phi'_3$ are 
the parity-mixed orbits, and their density show a triangle shape with the form
\begin{align}
\phi'^*_6({\bf r})  \phi'_6({\bf r}) &=\phi'^*_7({\bf r}) \phi'_7({\bf r})
\notag \\
&= \frac{1}{(\pi b^2)^{3/2}} \left( \frac{r}{b}\right)^2
e ^{-r^2/b^2} \left( 1+\frac{d}{2\sqrt{2}}\frac{r}{b}\cos(3\phi) +O(d^2) \right).
\end{align}

Similarly to the particle-hole 
representation of the $7\alpha$ wave function, 
$\Phi^{\rm BB}_{3\alpha}(\nu, d)$ of the order $d$ can be written in the particle-hole 
representation by using alternative definitions
$k\equiv2$, $\kb\equiv1$, and  $|0 \rangle_{\rm F}\equiv\Phi^{(0\hbar\omega)}_{3\alpha}(\nu, d)$,
\begin{equation}
\Phi^{\rm BB}_{3\alpha}(\nu, d)\approx
\prod_{\chi}\left(1-\frac{d}{2} a^\dagger_{-k,\chi} b^\dagger_{-\kb,-\chi} \right) 
\left(1+\frac{d}{2} a^\dagger_{+k,\chi} b^\dagger_{+\kb,-\chi} 
\right)  |0 \rangle_{\rm F}. 
\label{eq:3alpha-1p1h}
\end{equation}


\subsection{Development of the $7\alpha$ and $3\alpha$ states}
We calculate energies of 
the $7\alpha$ pentagon state $\Phi^{\rm BB}_{7\alpha}(\nu, d)$ and the $3\alpha$ triangle state
$\Phi^{\rm BB}_{3\alpha}(\nu, d)$ as functions of $\nu$ and $d$. 
The energies are evaluated by calculating expectation values of the Hamiltonian $H_{\rm eff}$
given in \ref{subsec:AMD} with respect to 
the intrinsic state $\Phi^{\rm BB}_{7\alpha}(\nu, d)$, and also the positive-parity projected state 
$(1+P_r)\Phi^{\rm BB}_{7\alpha}(\nu, d)$, the negative-parity projected state 
$(1-P_r)\Phi^{\rm BB}_{7\alpha}(\nu, d)$, and the $K^\pi=0^+$ projected one  $P^{K=0}(1+P_r)\Phi^{\rm BB}_{7\alpha}(\nu, d)$.
The parity projection corresponds to the restoration of the broken parity symmetry of the
intrinsic state, and the $K=0$ projection restores the broken axial symmetry.
The energy minimum state of $P^{K=0}(1+P_r)\Phi^{\rm BB}_{7\alpha}(\nu, d)$ may relate to 
the structure of the $K^\pi=0^+_1$ band in $^{28}$Si, while 
that of $(1-P_r)\Phi^{\rm BB}_{7\alpha}(\nu, d)$ corresponds to the $K^\pi=5^-_1$ band because 
$(1-P_r)\Phi^{\rm BB}_{7\alpha}(\nu, d)$ is equivalent to $K^\pi=5^-$ projected state at least 
in case of small $d$.
Note that the projected states may contain higher correlations beyond a 
single Slater determinant. 

\begin{figure}[th]
\epsfxsize=7 cm
\centerline{\epsffile{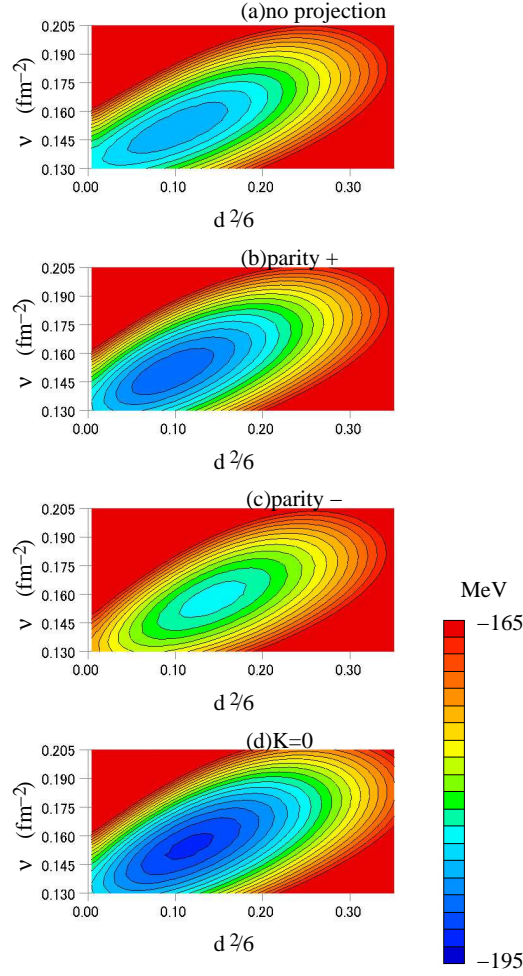}}
\caption{(a) Energy expectation values of $\Phi^{\rm BB}_{7\alpha}(\nu,d)$ plotted as functions of 
 $d^2/6$ and $\nu$. (b), (c), and (d) Those of the positive-parity state, the negative-parity state, and
 the $K=0$ state projected from the intrinsic wave function
$\Phi^{\rm BB}_{7\alpha}(\nu,d)$.
The parameter $d'$ is taken to be a small value, $d'/\sqrt{\nu}=0.1$ fm.
\label{fig:si28-nu-z2}
}
\end{figure}

The contour plots of energy surfaces are shown in Fig.~\ref{fig:si28-nu-z2} 
as functions of $\nu$ and $d^2/6$. 
It is found that the energy minimum of the energy surface for $\Phi^{\rm BB}_{7\alpha}(\nu, d)$
with no projection locates at $d^2/6\sim 0.1$. 
The finite pentagon size $d$ of the energy minimum indicates the development of pentagon shape,
namely, the spontaneous breaking of axial symmetry in the intrinsic structure.  
As seen in the minima of the energy surfaces shown in Fig.~\ref{fig:si28-nu-z2}(d) and (c),
the development of the pentagon shape enhances in the $K^\pi=0^+$ projected state, and it
is largest in the negative-parity projected state.

The value $d^2/6$ indicates approximately 
the mixing amplitude of the negative-parity component in the pentagon orbits, $\phi'_6$ and $\phi'_7$, as
given in Eq.~(\ref{eq:7alpha-d-expansion}).
The value $d^2/6\sim 0.1$ at the energy minimum of the positive-parity projected state 
indicates $\sim$10\% mixing which is comparable
to 5\% mixing of the negative-parity component in the pentagon orbits in the AMD wave function
$\Phi_{\rm AMD}({\bf Z}^{+})$ for $^{28}$Si.
The main reason for the smaller mixing in the AMD result than that in the ideal $7\alpha$ cluster model 
may be $\alpha$ cluster dissociation effects in the AMD calculations.

We also calculate energy of 
the equilateral triangle $3\alpha$ state $\Phi^{\rm BB}_{3\alpha}(\nu, d)$ 
as functions of $\nu$ and $d$.
Energies are calculated with respect to 
the intrinsic state $\Phi^{\rm BB}_{3\alpha}(\nu, d)$, and the positive-parity projected state 
$(1+P_r)\Phi^{\rm BB}_{3\alpha}(\nu, d)$, the negative-parity projected state 
$(1-P_r)\Phi^{\rm BB}_{3\alpha}(\nu, d)$, and the $K^\pi=0^+$ projected one $P^{K=0}(1+P_r)\Phi^{\rm BB}_{3\alpha}(\nu, d)$.
The energy minimum state of $P^{K=0}(1+P_r)\Phi^{\rm BB}_{3\alpha}(\nu, d)$ describes
the structure of the $K^\pi=0^+_1$ band in $^{12}$C, while 
that of $(1-P_r)\Phi^{\rm BB}_{3\alpha}(\nu, d)$ corresponds to the $K^\pi=3^-_1$ band. 

\begin{figure}[th]
\epsfxsize=7 cm
\centerline{\epsffile{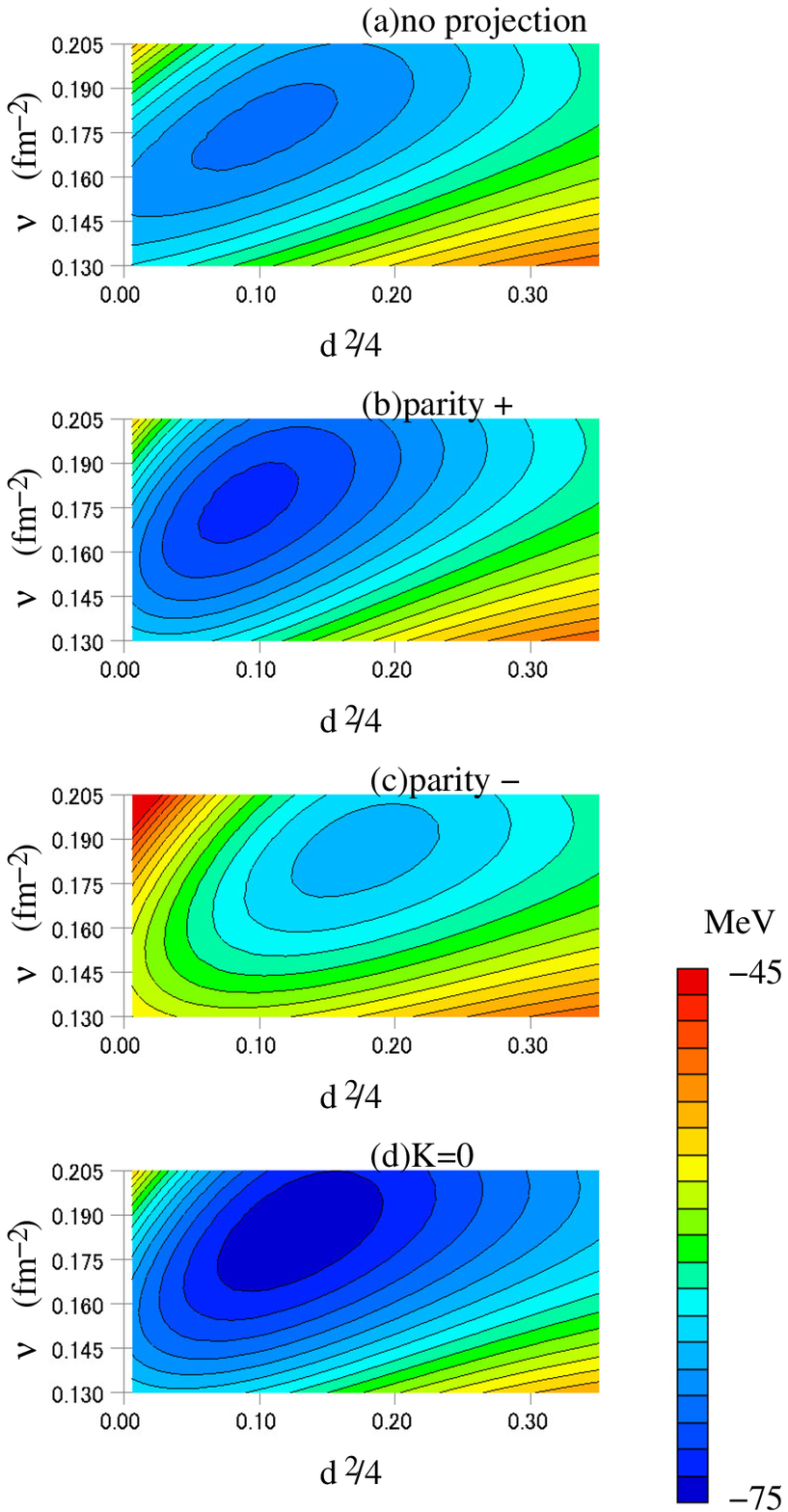}}
\caption{(a) Energy expectation values of $\Phi^{\rm BB}_{3\alpha}(\nu,d)$ plotted as functions of 
 $d^2/4$ and $\nu$. (b), (c), and (d) Those of the positive-parity state, the negative-parity state, and
 the $K=0$ state projected from the intrinsic wave function
$\Phi^{\rm BB}_{3\alpha}(\nu,d)$.
The parameter $d'$ is taken to be a small value, $d'/\sqrt{\nu}=0.1$ fm.
\label{fig:c12-nu-z2}
}
\end{figure}

The contour plots of the energy surfaces are shown in Fig.~\ref{fig:c12-nu-z2} 
as functions of $\nu$ and $d^2/4$. $d^2/4$ is approximately the mixing amplitude of 
$\phi_{0,0,\mp 2}$ in $\phi_{0,0,\pm 1}$ as described in Eq.~(\ref{eq:3alpha-d-expansion}).
The finite $d^2/4$ value of the energy minimum 
indicates the development of $3\alpha$ cluster structure. 
The energy minimum of the energy surface for $\Phi^{\rm BB}_{3\alpha}(\nu, d)$
with no projection locates at $d^2/4\sim 0.1$. 
The cluster development slightly enhances in the $K^\pi=0^+$ projected state, and 
it is most remarkable in the negative-parity projected state. 
This is consistent with the AMD calculations of $^{12}$C shown in Fig.~\ref{fig:dense}. 
One of the interesting features of $\Phi^{\rm BB}_{3\alpha}(\nu, d)$ 
is that the energy surface is quite shallow against the large triangle size $d$
at $\nu\sim 0.20$. This corresponds to three $\alpha$ cluster break up.

\subsection{Roles of the parity and $K^\pi$ projections in a mean-field picture}

As mentioned before, in the particle-hole representation 
based on $|\Phi^{(0\hbar\omega}_{7\alpha}\rangle=|0 \rangle_{\rm F}$, 
$\Phi^{\rm BB}_{7\alpha}(\nu, d)$ can be approximately expressed as Eq.~(\ref{eq:7alpha-1p1h}).
At least in the oder $d$, 
the negative-parity projected state is nothing but a linear combination of 
$1p$-$1h$ states,  
\begin{equation}
(1-P_r)|\Phi^{\rm BB}_{7\alpha}(\nu, d)\rangle \approx 
\frac{d}{\sqrt{6}} \sum_{\chi}\left( a^\dagger_{-k,\chi} b^\dagger_{-\kb,-\chi} 
- a^\dagger_{+k,\chi} b^\dagger_{+\kb,-\chi} \right) |0 \rangle_{\rm F}.
\label{eq:7alpha-1p1h-negative}
\end{equation}
Thus, $(1-P_r)\Phi^{\rm BB}_{7\alpha}(\nu, d)$ is described by the coherent sum of the $1p$-$1h$ states,
$a^\dagger_{-k,\chi} b^\dagger_{-\kb,-\chi}  |0 \rangle_{\rm F}$ and
$a^\dagger_{+k,\chi} b^\dagger_{+\kb,-\chi}  |0 \rangle_{\rm F}$.
It indicates that the negative-parity state can be described by the $K^\pi=5^-$ 
vibration mode on the oblate shape.
This is an alternative interpretation of the $K^\pi=5^-$ band. However, as already discussed before, 
since the axial symmetry of the oblate state is already broken in the intrinsic state before 
the negative-parity projection (see Fig.~\ref{fig:si28-nu-z2}), the $K^\pi=5^-$ band is regarded as the 
static pentagon ``shape'' instead of the $K^\pi=5^-$ vibration on the oblate state.

The positive-parity projected state corresponds to the mixing of $2p$-$2h$ states 
having $K=0$ and $K=\pm 10$ into the dominant 
$\Phi^{(0\hbar\omega)}_{7\alpha}$ state.
The high $K$ components do not affect to the $K^\pi=0^+$ ground band, and actually, 
they are dropped off in the $K=0$ projection.
Consequently, the $K^\pi=0^+$ projected state contains only the $2p$-$2h$ states with $K=0$, 
\begin{equation}
P^{K=0}(1+P_r)|\Phi^{\rm BB}_{7\alpha}(\nu, d)\rangle \approx |0 \rangle_{\rm F} 
- \frac{d^2}{6} \sum_{\chi}\sum_{\chi'} 
 a^\dagger_{-k,\chi} b^\dagger_{-\kb,-\chi} a^\dagger_{+k,\chi'} 
 b^\dagger_{+\kb,-\chi'}
|0 \rangle_{\rm F}.\label{eq:7alpha-1p1h-k0}
\end{equation}
The contained $2p$-$2h$ states are $k$ and $-k$ particle pairs and $\kb$ and $-\kb$ hole pairs, and they 
are associated with Cooper pairs in BCS theory~\cite{BCS}. 
Let us remind the reader that the normal pairings in nuclear systems 
are considered to be neutron-neutron pairing and proton-proton one in the spin $S=0$ channel. 
However,  the $2p$-$2h$ terms in Eq.~(\ref{eq:7alpha-1p1h-k0}) have not only spin-zero $nn$ and $pp$ pairs but also 
spin-one $np$ pairs. Namely, a $(S,T)=(1,0)$ (spin-one isoscalar) particle-particle pair and a
$(S,T)=(1,0)$  hole-hole pair couple to be totally $S=0$ and $T=0$, while 
a $(S,T)=(0,1)$ (spin-zero isovector) particle-particle pair and 
a $(S,T)=(0,1)$ hole-hole pair couple to be $S=0$ and $T=0$.
Because of the large number of coherent pairs, the $K^\pi=0^+$ state projected from the $7\alpha$ cluster state may gain much correlation energy.

In these analyses, we can say that, in both the negative-parity and $K^\pi=0^+$ states, 
the coherent particle-hole configurations 
due to the coherent edge DWs 
of four kinds $\chi=$ $p\uparrow$, $p\downarrow$, $n\uparrow$, and $n\downarrow$
play an important role in 
the development of the pentagon shape. We will show the importance of the coherence for the SSB in the later sections.

\section{Extension of BB $\alpha$-cluster models}\label{sec:analysis2}

As discussed in the previous section, the coherent proton and neutron edge DWs are essential to develop 
the pentagon and triangle shapes.
In this section, we investigate the development of the pentagon and triangle shapes without the proton-neutron coherence
by considering a pentagon or triangle neutron structure with a frozen  proton structure.
For this aim, we extend the BB $\alpha$-cluster wave functions
for the pentagon $7\alpha$- and the triangle $3\alpha$-cluster states as follows:
We assume the pentagon configurations of proton and neutron structures for a $Z=N=14$ system but take 
the pentagon size $d$ in Eq.~(\ref{eq:BB-penta1}) independently for protons and neutrons.
We take an enough small value of the pentagon size $d_p$ for protons and vary the size $d_n$ for neutrons.
Thus, the defined wave function can be written in the expansion of the order $d_n$ as 
\begin{align}
\Phi_{7\alpha\text{-}n}(\nu,d_n) &\approx
n'_0 \prod_{\sigma}
\det\{\phi'^{(0)}_1 \Chi_{p\sigma}, \cdots, \phi'^{(0)}_7\Chi_{p\sigma}
\} \notag\\
&\quad\times\prod_{\sigma} 
\det\{\phi'^{(0)}_1 \Chi_{n\sigma}, \cdots, \phi'^{(0)}_5\Chi_{n\sigma}, \phi'_6\Chi_{n\sigma},\phi'_7\Chi_{n\sigma} \},\notag\\
\phi'_6&=\phi_{(0,0,-2)} - \frac{d_n}{\sqrt{6}} \phi_{(0,0,+3)} + O(d_n^2), \notag\\
\phi'_7&=\phi_{(0,0,+2)} + \frac{d_n}{\sqrt{6}} \phi_{(0,0,-3)} + O(d_n^2).
\end{align}

In a similar way, we  also assume the triangle configurations of proton and neutron structures 
for a $Z=N=6$ system
by taking the triangle size $d$ in Eq.~(\ref{eq:BB-triangle1}) independently for protons and neutrons.
We take an enough small value of the triangle size $d_p$ for protons and vary the size $d_n$ for neutrons.
Then the wave function can be written in the expansion of the order $d_n$ as 
\begin{align}
\Phi_{3\alpha\text{-}n}(\nu,d_n) &\approx
n'_0 \prod_{\sigma}
\det\{\phi'^{(0)}_1 \Chi_{p\sigma} \phi'^{(0)}_2\Chi_{p\sigma} \phi'^{(0)}_3\Chi_{p\sigma}
\} \notag\\
&\quad\times\prod_{\sigma} 
\det\{\phi'^{(0)}_1 \phi'_2\Chi_{n\sigma}\phi'_3\Chi_{n\sigma} \},\notag\\
\phi'_2&=\phi'^{(0)}_2 + \frac{d_n}{2} \phi_{(0,0,+2)} + O(d_n^2), \notag\\
\phi'_3&=\phi'^{(0)}_3 - \frac{d_n}{2} \phi_{(0,0,-2)} + O(d_n^2). 
\end{align}

Moreover, we consider the triangle proton structure in a $Z=6$ and $N=14$ system 
to study the proton edge DW in a neutron-rich system. 
For the frozen neutron structure, we adopt a pentagon configuration of the neutron part 
with an enough small pentagon size $d_n$. The proton structure is assumed to be 
a triangle structure with the triangle size $d_p$, which is a variational parameter.
The wave function can be written in the expansion of the order $d_p$ as 
\begin{align}
\Phi_{^{20}\text{C-}p}(\nu,d_p) &\approx 
n'_0 \prod_{\sigma=\{\uparrow,\downarrow\}}
\det\{\phi'^{(0)}_1\Chi_{p\sigma}, \phi'_2\Chi_{p\sigma},\phi'_3\Chi_{p\sigma} \}\notag\\
&\quad\times\prod_{\sigma=\{\uparrow,\downarrow\}} 
\det\{\phi'^{(0)}_1 \Chi_{n\sigma}, \cdots, \phi'^{(0)}_7\Chi_{n\sigma}\}, \notag\\
\phi'_2&=\phi'^{(0)}_2 + \frac{d_p}{2} \phi_{(0,0,+2)} + O(d_p^2), \notag\\
\phi'_3&=\phi'^{(0)}_3 - \frac{d_p}{2} \phi_{(0,0,-2)} + O(d_p^2). 
\end{align}
This model corresponds to a $3\alpha$ core structure in $^{20}$C system. 

We calculate energies of $\Phi_{7\alpha\text{-}n}(\nu,d_n)$, $\Phi_{3\alpha\text{-}n}(\nu,d_n)$, 
and $\Phi_{^{20}\text{C-}p}(\nu,d_p)$ states and compare the results with 
$\Phi^{\rm BB}_{7\alpha}(\nu,d)$ and $\Phi^{\rm BB}_{3\alpha}(\nu,d)$. 
The energies are evaluated by calculating expectation values of the effective Hamiltonian $H_{\rm eff}$
for these states with no projection, the positive- and negative-parity 
projected states, and the $K^\pi=0^+$ projected states. 

\begin{figure}[th]
\begin{center}
\epsfxsize=5.5 cm
\epsffile{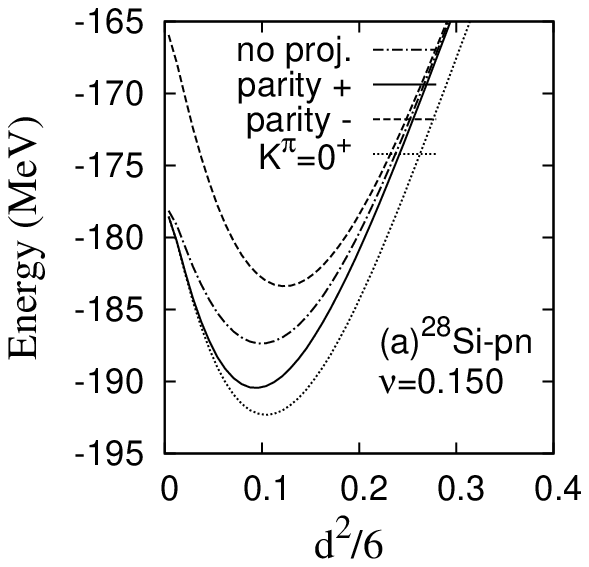}\\
\epsfxsize=5.5 cm
\epsffile{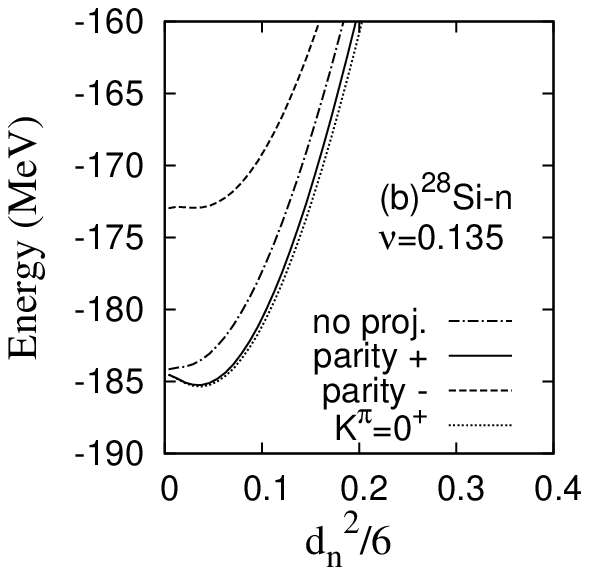}\\
\end{center}
\caption{
\label{fig:si28-z2}
(a) The energy of the $7\alpha$ state $\Phi^{\rm BB}_{7\alpha}(\nu,d)$ 
as a function of $d^2/6$. $d$ is the pentagon size for protons and neutrons.
(b) The energy of the $Z=N=14$ state $\Phi_{7\alpha\text{-}n}(\nu,d_n)$ 
with the frozen proton structure
as a function of $d_n^2/6$. $d_n$ is the pentagon size for neutrons. 
In each system, $\nu$ is fixed to be the optimum value at the energy minimum solution 
in the $\nu$-$d$ plane for the positive-parity projected states
as (a) $\nu=0.15$ fm$^{-2}$ 
 and (b) $\nu=0.135$ fm$^{-2}$. 
The parameter $d'$ is chosen to be $d'/\sqrt{\nu}=0.1$ fm.
}
\end{figure}

At first, we compare the pentagon size dependence of the energies of $\Phi_{7\alpha\text{-}n}(\nu,d_n)$ having the
frozen proton structure with that of $\Phi^{\rm BB}_{7\alpha}(\nu,d)$ having 
the proton-neutron coherent pentagon shapes. The energy curves are shown in Fig.~\ref{fig:si28-z2}. 
In each system, $\nu$ is fixed to be the optimum value 
at the energy minimum solution in the $\nu$-$d$ plane for the positive-parity projected state.
As already discussed in the previous section, $\Phi^{\rm BB}_{7\alpha}(\nu,d)$ shows the deep energy pocket 
around the energy minimum at the finite $d$ value [see Fig.~\ref{fig:si28-z2}(a)]. 
This indicates the
development of the pentagon shape, which corresponds to 
the spontaneous breaking of the axial symmetry of the oblate state $\Phi^{(0\hbar\omega)}_{7\alpha}$.
The potential pockets are deeper in the projected states than the intrinsic state with no projection.
In contrast to the energy curve for $\Phi^{\rm BB}_{7\alpha}(\nu,d)$, 
the energy curve for $\Phi_{7\alpha\text{-}n}(\nu,d_n)$ with no projection has the minimum around
$d_n=0$, which corresponds to the axial symmetric oblate state $\Phi^{(0\hbar\omega)}_{7\alpha}$. 
Even in the projected states, there is no deep pocket in a finite $d_n$ region, and 
the pentagon shape of the neutron structure is suppressed in the frozen proton structure.
This means that the neutron edge DW on the oblate state $\Phi^{(0\hbar\omega)}_{7\alpha}$
does not occur without the coherent proton edge DW. 
It may lead to 
suppression of pentagon shape in $Z\ne N$ nuclei.

\begin{figure}[th]
\begin{center}
\epsfxsize=5.5 cm
\epsffile{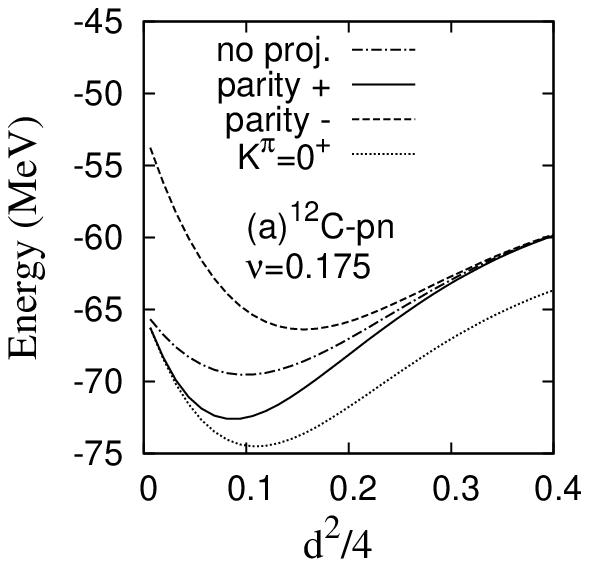}\\
\epsfxsize=5.5 cm
\epsffile{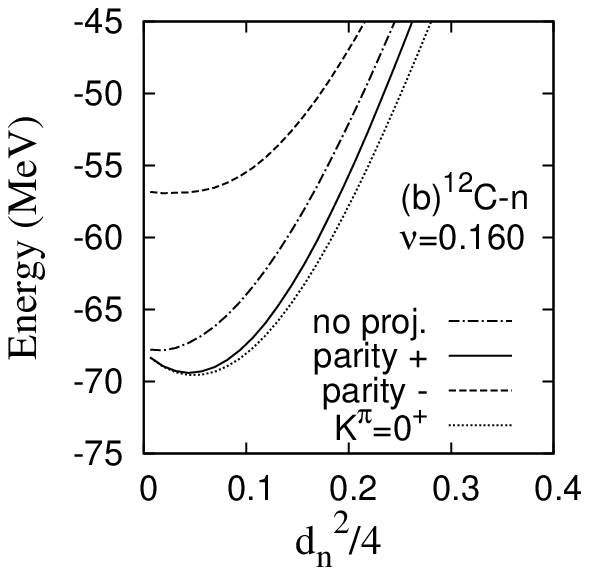}\\
\epsfxsize=5.5 cm
\epsffile{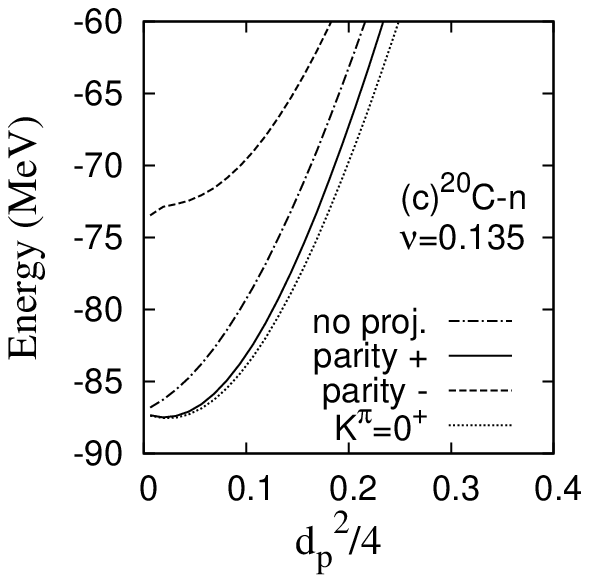}
\end{center}
\caption{
\label{fig:c12-z2}
(a) The energy of the $3\alpha$ state $\Phi^{\rm BB}_{3\alpha}(\nu,d)$
as a function of $d^2/4$, where $d$ is the pentagon size for protons and neutrons.
(b) The energy of the $Z=N=6$ state $\Phi_{3\alpha\text{-}n}(\nu,d_n)$ with the frozen proton structure
as a function of $d_n^2/4$. $d_n$ is the pentagon size for neutrons. 
(c) The energy of the $Z=6,N=14$ system $\Phi_{^{20}\text{C-}p}(\nu,d_p)$ 
with the frozen proton structure.
The pentagon size $d_n$ for the frozen neutron structure is taken to be $d_n^2=0.025$. 
In each system, $\nu$ is fixed to be the optimum value
 at the energy minimum solution
in the $\nu$-$d$ plane  for the positive-parity projected states
as  (a) $\nu=0.175$ fm$^{-2}$, 
 (b) $\nu=0.160$ fm$^{-2}$, and  
 (c) $\nu=0.135$ fm$^{-2}$.
}
\end{figure}

Next we discuss the triangle structures in $Z=6$ nuclei.
In a similar way to the pentagon structure, 
we compare triangle size dependence of the energies of $\Phi_{3\alpha\text{-}n}(\nu,d_n)$ having the
frozen proton structure with that of $\Phi^{\rm BB}_{3\alpha}(\nu,d)$ having 
the proton-neutron coherent triangle shapes in Figs.~\ref{fig:c12-z2}(a) and \ref{fig:c12-z2}(b). 
Again we find that each energy curve for $\Phi_{3\alpha\text{-}n}(\nu,d_n)$ has the minimum around
$d_n=0$, which corresponds to $\Phi^{(0\hbar\omega)}_{3\alpha}$. It is in contrast 
to the features of $\Phi^{\rm BB}_{3\alpha}(\nu,d)$, which shows the deep energy pocket 
at the finite $d$ indicating to the developed   
triangle shape.
We also consider the proton triangle shape in the $Z=6,N=14$ system,
$\Phi_{^{20}\text{C-}p}(\nu,d_p)$. The finite $d_p$ corresponds to the development of the 
$3\alpha$ core structure in $^{20}$C system with the oblate proton and neutron structures.
The energy of $\Phi_{^{20}\text{C-}p}(\nu,d_p)$ is the smallest around $d=0$ and increases 
as the triangle size $d_p$ becomes large. Compared with the energy curve of
$\Phi_{3\alpha\text{-}n}(\nu,d_n)$, the triangle proton structure is significantly unfavored in 
the $Z=6,N=14$ system. 

Then we conclude that the proton-neutron coherence is essential in development of the
pentagon and triangle structures of the oblate states in $Z=N$ nuclei.
Needless to say, this is consistent with the cluster aspect of $Z=N$ nuclei.
In the oblate state of neutron-rich C, the triangle cluster structure is suppressed.
The first reason for the quenching of cluster structure is 
lack of the proton-neutron coherence. The second reason is the expanded 
level spacing of proton orbits in neutron-rich nuclei 
because protons are deeply bound due to excess neutrons. Since the energy cost for 
$1p$-$1h$ proton excitations increases in the neutron-rich system, the correlation energy due to the triangle structure 
may not be able to overcome the cost.
We shall discuss the details in the next section.

\section{Edge density wave and spontaneous symmetry breaking}\label{sec:discussion}

As already mentioned, 
the pentagon and triangle structures can be interpreted as the static edge DWs at the
surface of the oblate states, which is expected to connect with 
the spontaneous symmetry breaking (SSB) of rotational invariance around the symmetric axis.

In systems with strong interaction, 
we know various SSB phenomena such as nuclear BCS, chiral symmetry breaking, and color superconductivities.
These SSB occur both homogeneously and inhomogeneously.
In particular, inhomogeneous SSB phases are discussed in the framework of nuclear DWs,
chiral DWs,
and Fulde-Ferrel-Larkin-Ovchinnikov state in color superconducting phase
\cite{overhauser60,tamagaki76,takatsuka78,migdal78,Dautry:1979bk,Deryagin:1992rw,Shuster:1999tn,Park:1999bz,Alford:2000ze,Nakano:2004cd,Giannakis:2004pf,Fukushima:2006su,Nickel:2009ke,Kojo:2009ha,Carignano:2010ac,Fukushima:2010bq},
whose phase breaks translational invariance and the corresponding condensation operator depends on 
the spatial coordinates.
In more general, the SSB resulting in a spatially nonuniform 
vacuum relates to condensation operators with finite momenta.
When we understand nuclear matter DW as 
the instability of the Fermi surface, the condensation operator
is given in the form $a^\dagger_{k_F} b^\dagger_{k_F}$, which has the momentum $2k_F$ ($k_F$ is
the Fermi momentum).
In condensed matter physics, inhomogeneous phases with DW are discussed as charge DWs and 
spin DWs~\cite{CDW,SDW}. 

These phenomena in infinite systems are discussed in terms of the order parameters,
which are characterized by non-zero expectation values of certain operators.
In finite systems, however, 
the symmetry cannot be broken in the energy eigenstates, 
because the symmetry is restored even if it is 
broken in the intrinsic state.
Nevertheless, it is useful 
to discuss the SSB in the state before projection or restoration 
by analyzing expectation values of specific operators resemble to the condensation operators 
as done for the BCS phenomena in finite nuclei.
For the pentagon and triangle structures, the expressions in Eqs.~(\ref{eq:7alpha-1p1h})
and (\ref{eq:3alpha-1p1h}) are the form similar to the 
matter DW operators $a^\dagger_{k_F} b^\dagger_{k_F}$.

In this section, we describe the SSB for edge DWs 
by introducing a simplified model in Appendix \ref{app:hamiltonian-simple}, and discuss 
the development and suppression of the pentagon and triangle structures from the 
viewpoint of the edge DW.
In this model, $\Phi^{(0\hbar\omega)}_{7\alpha}$ is assumed to be the Hartree-Fock 
vacuum $|0\rangle_{\rm F}$, and the orbits 
$|\phi_{(0,0,\pm \kb)} \Chi_{\tau\sigma}\rangle$ 
and $|\phi_{(0,0,\pm k)} \Chi_{\tau\sigma}\rangle$ 
are considered to be active Hartree-Fock single-particle states.
It means that the model space is truncated within 
$\phi_{(0,0,\pm k)}\Chi_{\tau\sigma}$ for particle states and 
$\phi_{(0,0,\pm \kb)}\Chi_{\tau\sigma}$ for hole states. 
This is equivalent to the model of 8 particles for 16 states, 
which is a kind of half filled models.
As for the residual interaction, we assume a contact interaction 
and adopt $H_{\rm DW}$ defined in Eq.~(\ref{eq:HDW}).
Note that this model is applicable also to the $3\alpha$ oblate state
by replacing $\kb=2$ and $k=3$ for the $7\alpha$ state 
with $\kb=1$ and $k=2$.
Then, the Hamiltonian in the particle-hole representation can be written as
\begin{align}\label{eq:DWHamiltonian}
H&=H_0+H_1+H_{\rm DW},\notag\\
H_0&= {_{\rm F}\langle} 0|H|0\rangle_{\rm F},\notag\\
H_1&=\sum_{\chi} E_{k,\tau} a^\dagger_{+k,\chi} a_{+k,\chi}
+\sum_{\chi} E_{k,\tau} a^\dagger_{-k,\chi} a_{-k,\chi} \notag\\
&\quad-\sum_{\chi} E_{\kb,\tau} b^\dagger_{+\kb,\chi} b_{+\kb,\chi}
-\sum_{\chi} E_{k,\tau} b^\dagger_{-\kb,\chi} b_{-\kb,\chi}, \notag\\
H_{\rm DW}&=2\sum_{\chi, \chi'} G^{({ph})}_{\chi,\chi'} \notag\\
&\quad\times\Bigl[a^\dagger_{+k,\chi} b^\dagger_{+\kb,-\chi} b_{+\kb,-\chi'} a_{+k,\chi'}\notag \\
&\quad\quad+a^\dagger_{-k,\chi} b^\dagger_{-\kb,-\chi} b_{-\kb,-\chi'} a_{-k,\chi'}\Bigr].
\end{align}
Here $\chi=\tau\sigma$ and $-\chi=\tau-\sigma$.

We use an ansatz for the new vacuum of the edge DWs with the axial-symmetry breaking as 
\begin{equation}\label{eq:waveFunction}
|\Psi\rangle = \prod_{\chi} \Bigl(v_{\tau}+u_{\tau} a^\dagger_{+k,\chi} b^\dagger_{+\kb,-\chi}\Bigr)
\prod_{\chi} \Bigl(v^*_{\tau}-u^*_{\tau} a^\dagger_{-k,\chi} b^\dagger_{-\kb,-\chi}\Bigr) |0\rangle _{\rm F},
\end{equation}
with \begin{equation}
|v_{\tau}|^2 + |u_{\tau}|^2 =1,
\end{equation}
where $v_{\tau}$ and $u_{\tau}$ are variational parameters determined by the energy variation, and time reversal invariance is taken into account. 
Our ansatz Eq.~(\ref{eq:waveFunction}) has the same form as that obtained in the approximation that the quantum fluctuation 
of the particle-hole operators 
such as $a^\dagger_{\pm k,\chi} b^\dagger_{\pm \kb,-\chi}$ 
are omitted as shown 
in Appendix~\ref{sec:meanField}.
It is clear that $\Psi$ is equivalent to 
$\Phi^{\rm BB}_{7\alpha}(\nu,d)$ with $u_\tau=-d/\sqrt{6}$
in the order $d$ approximation given by Eq.~(\ref{eq:7alpha-1p1h})
(or $\Phi^{\rm BB}_{3\alpha}(\nu,d)$ with $u_\tau=d/2$).
In general, the coefficients $u_\tau$ and $v_\tau$ are complex.
The phase $\phi_{0}$ of $u_\tau/v_\tau$ corresponds to the constant shift of the
rotation angle $\phi\rightarrow \phi+\phi_0$ in the density oscillation $\cos(5\phi)$ in Eq.~(\ref{eq:cos5phi}).
Since the phase $\phi_{0}$ for the lowest-energy solution is isospin independent; hereafter, $u_\tau$ and $v_\tau$ are taken to be real quantities.
The expectation values for this vacuum $|\Psi\rangle$ are 
\begin{align}
\bigl\langle a^\dagger_{\pm k,\chi} a_{\pm k,\chi} \bigr\rangle &=
\bigl\langle b^\dagger_{\pm \kb,-\chi} b_{\pm \kb,-\chi} \bigr\rangle =u_{\tau} u_{\tau},  \notag \\
\bigl\langle a^\dagger_{\pm k,\chi} b^\dagger_{\pm \kb,-\chi} \bigr\rangle&= 
\bigl\langle b_{\pm \kb,-\chi} a_{\pm k,\chi}  \bigr\rangle 
 =\pm  u_{\tau}  v_{\tau}.
\end{align}
The normal state $|0\rangle_{\rm F}$ has $v_\tau=1$ and $u_\tau=0$, while
the SSB vacuum has a finite $\langle a^\dagger_{\pm k,\chi} 
b^\dagger_{\pm \kb,-\chi} \rangle$, i.e., the finite value of 
$u_\tau v_\tau$.

The values $v_\tau$ and $u_\tau$ are determined by minimizing the expectation value 
$\langle \Psi|H| \Psi\rangle$, 
\begin{align}
E&=\langle \Psi|H| \Psi\rangle=  H_0 + 2 E_{\rm corr}, \notag\\
E_{\rm corr} &= \sum_{\tau\sigma} (E_{k,\tau}-E_{\kb,\tau}) u^2_{\tau} 
+ 2 \sum_{\chi,\chi'} G^{({ph})}_{\chi,\chi'} u_{\tau} v_{\tau} u_{\tau'}  v_{\tau'}.
\end{align}

For $Z=N$ systems, when isospin dependences of single-particle energies $E_{k,\tau}$ and 
$E_{\kb,\tau}$ are ignored, 
$v_\tau$ and $u_{\tau}$ do not depend on the isospin $\tau$, and the energy correction 
$E_{\rm corr}$ from the energy $H_0$ is
\begin{align}\label{eq:ecorrvp}
E_{\rm corr} &= \sum_{\tau\sigma} (E_{k,\tau}-E_{\kb,\tau}) u^2_{\tau} 
+ 2 \sum_{\chi\ne\chi'} g^{({ph})} u_{\tau} v_{\tau} u_{\tau'}  v_{\tau'} \notag \\ 
&= 4 \{(E_{k}-E_{\kb}) u^2 + 6 g^{({ph})} u^2v^2 \}.
\end{align}
The stationary condition with respect to variations of $u$ and $v$ with the constraint 
$u\delta u+v\delta v=0$ leads to the equation
\begin{equation}
(E_{k}-E_{\kb})u-6g^{({ph})}u(u^2-v^2)=0.
\end{equation}
For a non-zero $u$, $u$ and $v$ are solved, 
\begin{align}\label{eq:theSolutionOfMeanField}
u^2&=\frac{1}{2}\left( 1+ \frac{E_{k}-E_{\kb}}{6g^{({ph})}}\right), \notag\\
v^2&=\frac{1}{2}\left( 1- \frac{E_{k}-E_{\kb}}{6g^{({ph})}}\right), \notag\\
uv&=\frac{1}{2}\sqrt{1- \left(\frac{E_{k}-E_{\kb}}{6g^{({ph})}}\right)^2 }.
\end{align}
It turns out that, to obtain a non-zero $uv$ with real $u$ and $v$ values for the SSB vacuum
the following condition must be satisfied:
\begin{equation}\label{eq:ssb-pn}
E_{k}-E_{\kb} < - 6g^{({ph})}.
\end{equation}
This indicates that the SSB occurs provided that the strength $-g^{({ph})}$ of the attraction is 
large enough so as to satisfy the above condition. In other words, the static edge DWs can exist
if the correlation energy $- 6g^{({ph})}$ overcomes the energy cost $E_{k}-E_{\kb}$ of a $1p$-$1h$ excitation. For $N=Z$ systems, one can regard the correlation of $1p$-$1h$  
as that of four particles,
because $1h$ corresponds to the three particles state in the particle picture.

Let us consider the role of the proton-neutron coherence in the SSB.
In the case that there is no proton-neutron interaction, 
the coupling $G^{({ph})}_{\chi,\chi'}$ is taken to be $G^{({ph})}_{\chi,\chi'}
=g^{({ph})}\delta_{\tau\tau'}(1-\delta_{\sigma\sigma'})$. 
Protons and neutrons are decoupled in the Hamiltonian, and the energy correction 
\begin{equation}\label{eq:ecorr-decouple}
E_{\rm corr} = 4 \{(E_{k}-E_{\kb}) u^2 + 2 g^{({ph})} u^2v^2 \}
\end{equation}
leads to the condition for the SSB,
\begin{equation}
E_{k}-E_{\kb} < - 2g^{({ph})}.
\end{equation}
This condition is more difficult to be satisfied than Eq.~(\ref{eq:ssb-pn}).
This is the reason why the proton and neutron coherent edge DWs can be stable, 
while the incoherent neutron or proton edge DW is unfavored in the oblate 
$7\alpha$ and $3\alpha$ states.
The reason for the three times smaller interaction term, i.e., the correlation energy in 
Eq.~(\ref{eq:ecorr-decouple}) than that in Eq.~(\ref{eq:ecorrvp})
is that, in the particle picture, 
$1h$ corresponds to the one particle state in case with no proton-neutron interaction,
instead of the $1h$ state corresponding to the three particles state
in case with proton-neutron interactions.

We also consider the further unfavored proton edge DW 
in a neutron-rich system discussed in the previous section. 
Protons are deeply bound in a neutron-rich system, and therefore the energy cost 
$E_{k}-E_{\kb}$ for the $1p-1h$ excitation becomes large in general.
As a result, the condition $E_{k}-E_{\kb} < - 2g^{({ph})}$ becomes severe, and 
the proton edge DW is suppressed largely in neutron-rich nuclei.


\section{Summary and outlook}\label{sec:summary}

Pentagon and triangle shapes in $^{28}$Si and $^{12}$C were discussed in the relation with 
DW at the edge of the oblate states.
In the AMD calculations,
the  $K^\pi=5^-$ band in $^{28}$Si and the $K^\pi=3^-$ band in $^{12}$C are described
by the pentagon and triangle shapes, respectively.
These negative-parity bands can be interpreted as the parity partners
of the  $K^\pi=0^+$ ground bands and they are constructed from the parity-asymmetric-intrinsic states.
The pentagon and triangle shapes originate in the $7\alpha$ and $3\alpha$ cluster structures.

We performed analysis of ideal cluster model wave functions 
using BB $\alpha$-cluster wave functions and also extended BB wave functions, 
and investigated the development of the pentagon and triangle shapes. 
It was found that 
the proton-neutron coherence is essential in development of the
pentagon and triangle structures of the oblate states in $Z=N$ nuclei.
Without the proton-neutron coherent density oscillation, the 
pentagon and triangle shapes are suppressed. 
Needless to say, this is consistent with the features of light $Z=N$ nuclei,
in which cluster structures are favored because of $\alpha$-cluster formation. 
In the oblate state of neutron-rich C, the triangle cluster structure is suppressed.

In analysis of single-particle orbits of the AMD wave functions and BB $\alpha$
cluster wave functions, 
the pentagon and triangle shapes are regarded as the static 
one-dimensional DWs at the edge of the oblate states.
The edge DWs can be described by nonuniform orbits with parity mixing, which give 
density oscillation with the wave number five and three at the surface of the 
$0\hbar\omega$ oblate states. 

The static edge DWs of the oblate $Z=N$ nuclei
are understood by the spontaneous symmetric breaking (SSB) 
of rotational invariance around the symmetric axis of the oblate states. 
In other words, the development of the $7\alpha$ and $3\alpha$ cluster structures
is interpreted as the instability of axial symmetry
with respect to the pentagon and the triangle shapes.
We introduced a simplified model and discussed the SSB for the edge DWs.
The development and the suppression of the pentagon 
and triangle structures are described by the SSB inducing 
static edge  DWs. 

In the simplified model, the $0\hbar\omega$ oblate states are assumed to be 
the Hartree-Fock vacuums $|0\rangle_{\rm F}$. 
The model space for particle and hole states are truncated so that only
$\phi_{(0,0,\pm 3)}\Chi_{\tau\sigma}$ and 
$\phi_{(0,0,\pm 2)}\Chi_{\tau\sigma}$ are active. 
Assuming a contact interaction, we adopted the DW term $H_{\rm DW}$
as the residual interaction. 
For the proton-neutron coherent edge DWs in $Z=N$ systems, 
the SSB occurs, when the condition 
$E_{k}-E_{\kb} < - 6g^{({ph})}$ is satisfied.
If there is no coupling between protons and neutrons, 
the condition for the SSB is $E_{k}-E_{\kb} < - 2g^{({ph})}$, which is more 
severe condition than the proton-neutron coherent case.
This means that the proton and neutron coherent edge DWs are favored,  
while an incoherent neutron or proton edge DW is unfavored.

Considering the condition $E_{k}-E_{\kb} < - 2g^{({ph})}$ for an incoherent edge DW, 
we explained the reasons why the triangle cluster structure is suppressed
in the oblate state of neutron-rich C.
Since protons are deeply bound in neutron-rich nuclei, the level spacing of proton orbits becomes large.
It increases the energy cost $E_{k}-E_{\kb}$ for 
a $1p$-$1h$ excitation, and hence, the correlation energy $- 2g^{({ph})}$ due to the triangle structure 
is not able to overcome the cost $E_{k}-E_{\kb}$.

The scenario for the suppression of the proton DW in neutron-rich systems could be
extended also to infinite matter problems. Let us mention about the instability with respect to 
 proton density oscillation in a neutron-rich matter, a symmetric nuclear matter, and 
a pure proton matter 
with the same Fermi momentum of protons ignoring the Coulomb force.
The proton density wave should be most unfavored in the neutron-rich matter among these three cases, 
while that might be favored with the coherent neutron DW in the symmetric nuclear matter.
It turns out that the possibility of $\alpha$-cluster crystallization in the neutron-rich matter
may be suspicious. Alternatively, we can say that the $\alpha$-cluster crystallization may be suppressed 
in the neutron-rich matter because of the quenched effective mass of protons.

In the present simplified model, in which active orbits are limited to be a small number, 
DW may be superior to BCS-type pairing in $Z=N$ systems. 
We adopted the ansatz of the residual interaction $H_2=H_{\rm DW}$ and discuss the
edge DWs in relation to the SSB. 
This ansatz may be applicable only to the case that 
the level density is enough low, active orbits are restricted in almost one dimension, and
the spin-orbit force can be ignored. 
The oblate $^{12}$C may satisfy this condition and the oblate $^{28}$Si would do probably.
However, we should comment that, in normal nuclei, static surface DWs may 
yield to the BCS pairing.
The spin-orbit force may also weaken static DWs.
Moreover, when the number of active orbits are large enough, the BCS pairing overcomes to 
DWs. Therefore, 
in heavy-mass nuclei, especially, in spherical nuclei, the BCS pairing can be 
predominant as well known.
In fact, various phenomena due to the BCS pairing
has been observed in heavy-mass nuclei and 
are successfully described by the BCS theory in the $j$-$j$ coupling scheme.

\appendix
\section{H.O. single-particle states}\label{appendix:ho}
To see the relation between BB cluster wave functions and shell model wave functions,
it is convenient to expand a BB wave function with $LS$-coupling shell model wave functions
which are described in terms of single-particle orbits in the spherical H.O. potential.
For instance, an $\alpha$ cluster located at the origin is expressed by four nucleons, 
$p\uparrow$, $p\downarrow$, $n\uparrow$, and $n\downarrow$ occupying the $0s$ orbit 
in the H.O. potential with the frequencies $\omega_x=\omega_y=\omega_z=\omega\equiv \hbar/mb^2$.
Here the parameter $b$ is related to $\nu$ of cluster wave functions as $\nu=1/2b^2$.
For oblate systems, it is convenient to use the expression of single-particle orbits 
with cylinder coordinates, $\rho=\sqrt{x^2+y^2},z,\phi$, where $z$ is the symmetry axis. 
Then, H.O. single-particle orbits are characterized by quantum numbers $n_z$, $n_\rho$, $m_l$. 
Here
$n_z$ and $n_\rho$ are the node numbers with respect to $z$ and $\rho$ coordinates, 
$m_l$ is the eigenvalue for the $z$-component of the orbital angular momentum. 
The total quantum number is $N=n_z+2n_\rho+|m_l|$.

The explicit forms of the H.O. single-particle orbits $\phi_{(n_z,n_\rho,m_l)}$ 
for $(n_z,n_\rho,m_l)$=$(0,0,\pm1)$, $(0,0,\pm 2)$, and $(0,0,\pm 3)$ are 
\begin{align}
\phi_{(0,0,\pm 1)}({\bf r})&=\frac{\mp 1}{(\pi b^2)^{3/4}} \frac{\rho}{b} e^{\pm i\phi} e ^{-r^2/2b^2},\notag\\
\phi_{(0,0,\pm 2)}({\bf r})&=\frac{1}{\sqrt{2}(\pi b^2)^{3/4}} \left( \frac{\rho}{b}\right)^2 e^{\pm 2i\phi} e ^{-r^2/2b^2},\notag\\
\phi_{(0,0,\pm 3)}({\bf r})&=\frac{\mp 1}{\sqrt{6}(\pi b^2)^{3/4}} \left( \frac{\rho}{b}\right)^3 e^{\pm 3i\phi} e ^{-r^2/2b^2}.
\end{align}

\section{Particle and hole representation}
In this appendix,  we summarize the notations of the particle and hole representation.
The creation and annihilation operators, $c_{\alpha}^\dagger$ and $c_{\alpha}$ for 
a state $|\alpha \rangle$ are defined as
\begin{align}
c_{\alpha}^\dagger |-\rangle &= | \alpha \rangle,\notag\\
c_{\alpha} |\alpha \rangle &= | - \rangle,\notag\\
c_{\alpha}^\dagger |\alpha \rangle &= 0,\notag\\
c_{\alpha} |- \rangle &= 0,
\end{align}
where
$|-\rangle$ is the no-particle state, and $\alpha$ denotes the index of all degrees of freedom of the single-particle state such as momentum, spin, and isospin.
  $c_\alpha$ and $c_\beta^\dag$ satisfies 
$\{c_\alpha,c_\beta^\dag\}=\delta_{\alpha,\beta}$, and other anticommutation relations are zero.
To describe particle-hole excitations on the Hartree-Fock (HF) vacuum, we define the HF vacuum state as
\begin{equation}
|0\rangle_F \equiv  \prod_{\alpha < F} c_{\alpha}^\dagger |-\rangle,
\end{equation}
and  the particle and hole operators as 
\begin{align}\label{eq:aAndb}
a_{\alpha}^\dagger &= c_{\alpha}^\dagger \qquad\qquad {\rm for} \quad \alpha > F, \notag\\
b_{\alpha}^\dagger &= S_{-\alpha} c_{-\alpha} \quad\;\; {\rm for} \quad \alpha < F, \notag\\
a_{\alpha} &= c_{\alpha} \qquad\qquad {\rm for} \quad \alpha > F,  \notag\\
b_{\alpha} &= S_{-\alpha} c_{-\alpha}^\dagger \quad\;\; {\rm for} \quad \alpha < F.
\end{align}
Here $\alpha < F$ and $\alpha > F$ means the states below and above the Fermi surface, respectively. 
The time reversal state of $|\alpha\rangle$ is defined as $S_{-\alpha}|-\alpha\rangle$.

For an infinite matter of spin-1/2 fermions, 
single-particle states can be characterized by momentum $k$ and spin $s_z=\sigma$.
In a usual convention,  $|-\alpha\rangle=|-k,-\sigma\rangle$ for $|\alpha\rangle=|k,\sigma \rangle$ and 
\begin{equation} \label{eq:alpha-matter}
S_{\alpha}\equiv (-1)^{\frac{1}{2}-\sigma_\alpha}.
\end{equation}
For a spherically symmetric system, single-particle states can be characterized by 
the quantum numbers $|\alpha\rangle\equiv |nlsjm_j \rangle$ in the $j$-$j$ coupling picture, 
and the corresponding $|-\alpha\rangle$ 
and the phase convention are
\begin{align}
|-\alpha\rangle&=|nlsj-m_j \rangle, \notag \\
S_{\alpha} &\equiv (-1)^{j -m_j}.
\end{align}

In an axial symmetric system in the $l$-$s$ coupling scheme
such as the present $7\alpha$ and $3\alpha$ models for $^{28}$Si and $^{12}$C,
we use the notation 
 $|\alpha\rangle=|n_z n_\rho m_l \sigma\rangle$ specified by the quantum numbers 
in the cylinder coordinates and adopt the following conventions:
\begin{align}\label{eq:alpha-axial}
|-\alpha\rangle&=|n_z n_\rho m_l -\sigma \rangle, \notag\\
S_{\alpha} &\equiv (-1)^{\frac{1}{2} -\sigma-m_l}.
\end{align}
The operator 
$b^\dagger_\alpha$ creates a hole carrying the $z$-component of angular momentum $m_l$
and the spin $s_z=\sigma$.  

We consider the Hamiltonian including the two-body interaction
\begin{align}
H &= \sum_{\alpha\beta} \langle\alpha|T|\beta\rangle c^\dagger_\alpha c_\beta + \frac{1}{2} 
\sum_{\alpha\beta\gamma\delta} {\cal V}_{\alpha,\beta,\gamma,\delta} c^\dagger_\alpha c^\dagger_\beta
c_\gamma c_\delta,\notag\\
{\cal V}_{\alpha,\beta,\gamma,\delta} &\equiv \frac{1}{2}  \{ \langle\alpha\beta |v| \gamma\delta\rangle -
\langle\alpha\beta |v| \delta\gamma\rangle \}.
\end{align}
We rewrite the Hamiltonian in normal-ordered form with respect to new particle and hole operators
assuming that the single-particle states, $\alpha$, are solutions of Hartree-Fock single-particle equations, 
which diagonalize the Hamiltonian matrix 
\begin{equation}
\langle\beta|T|\gamma\rangle = \sum_{\alpha<F}  [ \langle\alpha\beta |v| \alpha\delta\rangle -
\langle\alpha\beta |v| \delta\alpha\rangle ] =E_\beta \delta_{\beta\delta}.
\end{equation}
Then the Hamiltonian takes the form, 
\begin{equation}
H=H_0+H_1+ H_2,
\end{equation}
with 
\begin{align}\label{eq:hamiltonian}
H_0&=\sum_{\alpha<F}\langle\alpha|T|\alpha\rangle + \frac{1}{2} 
\sum_{\alpha<F}\sum_{\beta<F}
  [ \langle\alpha\beta |v| \alpha\beta\rangle -
\langle\alpha\beta |v| \beta\alpha\rangle ] , \notag\\
H_1&=\sum_{\alpha>F} E_\alpha a_\alpha^\dagger a_\alpha 
-\sum_{\alpha <F} E_\alpha b_\alpha^\dagger b_\alpha, \notag\\
H_2&= \frac{1}{2} \sum _{\alpha\beta\gamma\delta} 
{\cal V}_{\alpha,\beta,\gamma,\delta}N(c^\dagger_\alpha c^\dagger_\beta c_\delta c_\gamma),
\end{align}
where $N( \ \ )$ is the normal-ordered product with respect to the particle and hole operators defined before.
The residual interaction $H_2$ contains the
particle-particle, hole-hole, and particle-hole scattering, 
\begin{align}\label{eq:h2}
H_{{pp}}&=\frac{1}{2}\sum _{\alpha, \beta,\gamma,\delta>F} 
{\cal V}_{\alpha,\beta,\gamma,\delta} 
a^\dagger_\alpha a^\dagger_\beta a_\delta a_\gamma, \notag\\
H_{{hh}}&=\frac{1}{2}\sum _{\alpha, \beta,\gamma,\delta<F} 
{\cal V}_{\alpha,\beta,\gamma,\delta} 
b^\dagger_\alpha b^\dagger_\beta b_\delta b_\gamma,\notag\\
H_{{ph}}&=2\sum _{\alpha, \gamma>F} \sum _{\beta, \delta < F} 
{\cal V}_{\alpha,-\delta,-\beta,\gamma} S_{-\beta}S_{-\delta}
a^\dagger_\alpha b^\dagger_\beta b_\delta a_\gamma.
\end{align}

\section{Hamiltonian of the simplified model} \label{app:hamiltonian-simple}
We introduce a simplified model for the oblate state
$\Phi^{(0\hbar\omega)}_{7\alpha}$. In this model, 
$\Phi^{(0\hbar\omega)}_{7\alpha}$ is assumed to be the Hartree-Fock (HF)
vacuum $|0\rangle_{\rm F}$, and possible 
particle-hole excitations are restricted within the HF single-particle states of 
$|\phi_{(0,0,\pm k)} \Chi_{\tau\sigma}\rangle$ and $|\phi_{(0,0,\pm \kb)} \Chi_{\tau\sigma}\rangle$.
This means that the model space is truncated so that active orbits are only $|\phi_{(0,0,\pm k)} \Chi_{\tau\sigma}\rangle$
for particle states and $|\phi_{(0,0,\pm \kb)} \Chi_{\tau\sigma}\rangle$ for hole states
with $k=3$ and $\kb=2$ (or $k=2$ and $\kb=1$ for the $\Phi^{(0\hbar\omega)}_{3\alpha}$). 
We use the labels $\alpha=\pm k,\tau\sigma$ and $\alpha=\pm \kb,\tau\sigma$
for these active single-particle states and also adopt 
the notations $\chi\equiv\tau\sigma$ and $-\chi\equiv\tau-\sigma$.
In this paper, we define the particle and hole operators as  
\begin{align}
a_{\pm k,\chi}^\dagger &= c_{\pm k,\chi}^\dagger,\notag\\
b_{\pm \kb,\chi}^\dagger &= c_{\mp\kb,-\chi}.
\end{align}
Here, for convenience,  we adopt the definition of the hole operators without the phase convention instead of Eq.~(\ref{eq:aAndb}). 

In this model, we assume a contact two-body attraction  $v(r) =g\delta(r)$
with $g< 0$ for the residual interaction in the $H_2$ term.
The matrix element $V_{\alpha,\beta,\gamma,\delta}$ is not zero only when $k_\alpha+k_\beta=k_\gamma+k_\delta$ 
and $\chi_\alpha=\chi_\gamma\ne \chi_\beta=\chi_\delta$ (or $\chi_\alpha=\chi_\delta\ne \chi_\beta =\chi_\gamma$)
are satisfied and calculated to be
\begin{align}
G^{({pp})}_{\chi\chi'} & \equiv V_{k\chi,-k\chi', k\chi,-k\chi'}
= g^{({pp})} (1-\delta_{\chi\chi'}), \notag\\
 g^{({pp})} &\equiv \frac{g}{2}\langle k,-k| \delta(r) | k,-k \rangle,  \notag\\
G^{({hh})}_{\chi\chi'} & \equiv V_{{\kb}\chi,-{\kb}\chi',{\kb}\chi,-{\kb}\chi'}
=  g^{({hh})} (1-\delta_{\chi\chi'}), \notag\\
g^{({hh})} &\equiv  \frac{g}{2}\langle {\kb},-{\kb}| \delta(r) | {\kb},-{\kb} \rangle, \notag\\
G^{({ph})}_{\chi\chi'} & \equiv V_{ k\chi,-\kb\chi', k\chi,-\kb\chi'}
= g^{({ph})}(1-\delta_{\chi\chi'}), \notag\\
g^{({ph})} &\equiv  \frac{g}{2}\langle k,-\kb| \delta(r) | k,-\kb \rangle.
\end{align}
Using the symmetry (or antisymmetry) of the matrix elements $V_{\alpha,\beta,\gamma,\delta}$ with respect to indexes, 
the Hamiltonian Eq.~(\ref{eq:hamiltonian}) in the particle-hole representation can be rewritten in the explicit form, 
\begin{align}
H&=H_0+H_1+H_2, \notag\\
H_0&=  {_{\rm F}\langle} 0|H|0\rangle_{\rm F}, \notag\\
H_1&=\sum_{\chi} E_{k,\chi} a^\dagger_{+k,\chi} a_{+k,\chi}
+\sum_{\chi} E_{k,\chi} a^\dagger_{-k,\chi} a_{-k,\chi} \notag \\
&\quad-\sum_{\chi} E_{\kb,\chi} b^\dagger_{+\kb,\chi} b_{+\kb,\chi}
-\sum_{\chi} E_{k,\chi} b^\dagger_{-\kb,\chi} b_{-\kb,\chi}, \notag \\
H_2&= H^{{ph}}+H^{{pp}}+H^{{hh}},
\end{align}
with
\begin{align}
H^{{ph}}&=2\sum_{\chi,\chi'} G^{({ph})}_{\chi,\chi'} 
\Bigl[a^\dagger_{+k,\chi} b^\dagger_{+\kb,-\chi} b_{+\kb,-\chi'} a_{+k,\chi'} \notag\\
&\quad\quad+a^\dagger_{-k,\chi} b^\dagger_{-\kb,-\chi} b_{-\kb,-\chi'} a_{-k,\chi'} \notag\\
&\quad\quad+a^\dagger_{+k,\chi} b^\dagger_{-\kb,-\chi} b_{-\kb,-\chi'} a_{+k,\chi'} \notag\\
&\quad\quad+a^\dagger_{-k,\chi} b^\dagger_{+\kb,-\chi} b_{+\kb,-\chi'} a_{-k,\chi'}\Bigr] \notag\\
&\quad-2\sum_{\chi, \chi'} G^{({ph})}_{\chi,\chi'}
\Bigl[a^\dagger_{+k,\chi} b^\dagger_{+\kb,-\chi'} b_{+\kb,-\chi'} a_{+k,\chi} \notag\\
&\quad\quad+a^\dagger_{-k,\chi} b^\dagger_{-\kb,-\chi'} b_{-\kb,-\chi'} a_{-k,\chi} \notag\\
&\quad\quad+a^\dagger_{+k,\chi} b^\dagger_{-\kb,-\chi'} b_{-\kb,-\chi'} a_{+k,\chi} \notag\\
&\quad\quad+a^\dagger_{-k,\chi} b^\dagger_{+\kb,-\chi'} b_{+\kb,-\chi'} a_{-k,\chi}\Bigr],\\
H^{{pp}}
&=\sum_{\chi, \chi'} G^{({pp})}_{\chi,\chi'}\Bigl[a^\dagger_{+k,\chi} a^\dagger_{+k,\chi'} a_{+k,\chi'} a_{+k,\chi} \notag\\
&\quad\quad+a^\dagger_{-k,\chi} a^\dagger_{-k,\chi'} a_{-k,\chi'} a_{-k,\chi} \Bigr]\notag\\
&\quad\quad+2\sum_{\chi, \chi'} G^{({pp})}_{\chi,\chi'}\Bigl[a^\dagger_{+k,\chi} a^\dagger_{-k,\chi'} a_{-k,\chi'} a_{+k,\chi}\notag\\
&\quad\quad+a^\dagger_{+k,\chi} a^\dagger_{-k,\chi'} a_{+k,\chi'} a_{-k,\chi}\Bigr],\\
H^{{hh}}
&=\sum_{\chi, \chi'} G^{({hh})}_{\chi,\chi'}\Bigl[b^\dagger_{+{\kb},\chi} b^\dagger_{+{\kb},\chi'} b_{+{\kb},\chi'} b_{+{\kb},\chi}\notag \\
&\quad\quad+b^\dagger_{-{\kb},\chi} b^\dagger_{-{\kb},\chi'} b_{-{\kb},\chi'} b_{-{\kb},\chi} \Bigr]\notag\\
&\quad\quad+2\sum_{\chi,\chi'} G^{({hh})}_{\chi,\chi'}\Bigl[b^\dagger_{+{\kb},\chi} b^\dagger_{-{\kb},\chi'} b_{-{\kb},\chi'} b_{+{\kb},\chi}\notag\\
&\quad\quad+b^\dagger_{+{\kb},\chi} b^\dagger_{-{\kb},\chi'} b_{+{\kb},\chi'} b_{-{\kb},\chi}\Bigr],
\end{align}
where we omit $0\to4$ and $1\to3$ and their inverse processes.
In the particle-hole interaction term $H^{{ph}}$, the DW term
\begin{align}\label{eq:HDW}
H_{\rm DW} &\equiv 2\sum_{\chi, \chi'} G^{({ph})}_{\chi,\chi'} 
\Bigl[a^\dagger_{+k,\chi} b^\dagger_{+\kb,-\chi} b_{+\kb,-\chi'} a_{+k,\chi'} \notag\\
&\quad\quad+a^\dagger_{-k,\chi} b^\dagger_{-\kb,-\chi} b_{-\kb,-\chi'} a_{-k,\chi'}\Bigr]
\end{align}
may induce the edge DWs having the wave number $\pm(k+\kb)$, which 
gives the non-zero expectation value $\langle a^\dagger_{\pm k,\chi} b^\dagger_{\pm \kb,-\chi}\rangle$.
The terms of $a^\dagger_{\pm k,\chi} b^\dagger_{\mp \kb,-\chi} b_{\mp \kb,-\chi'} a_{\pm k,\chi'}$ in $H^{{ph}}$
may induce the exciton mode having the wave number $\pm 1$. They contains the spurious mode
of  translational motion and are of no interest in finite systems.
Other terms in $H^{{ph}}$ have the opposite sign and they do not give coherent effects to the correlation energy.

In $H^{{pp}}$ and $H^{{hh}}$, the interactions which may induce the BCS pairing are the following terms:
\begin{align}
H^{{pp}}_{\rm BCS}
&=2\sum_{\chi, \chi'} G^{({pp})}_{\chi,\chi'}\Bigl[a^\dagger_{+k,\chi} a^\dagger_{-k,\chi'} a_{-k,\chi'} a_{+k,\chi}\\
&\quad+a^\dagger_{+k,\chi} a^\dagger_{-k,\chi'} a_{+k,\chi'} a_{-k,\chi}\Bigr],\notag\\
H^{{hh}}_{\rm BCS}
&=2\sum_{\chi,\chi'} G^{({hh})}_{\chi,\chi'}\Bigl[b^\dagger_{+{\kb},\chi} b^\dagger_{-{\kb},\chi'} b_{-{\kb},\chi'} b_{+{\kb},\chi}\notag\\
&\quad+b^\dagger_{+{\kb},\chi} b^\dagger_{-{\kb},\chi'} b_{+{\kb},\chi'} b_{-{\kb},\chi}\Bigr].
\end{align}
In the case of $Z=N$ nuclei, only two types of BCS pairing, for instance,  $a^\dagger_{+k,p\uparrow} a^\dagger_{-k,p\downarrow}$ and 
$a^\dagger_{+k,n\uparrow} a^\dagger_{-k,n\downarrow}$, are usually considered
among four species of nucleons, $\chi=p\uparrow,p\downarrow,n\uparrow,n\downarrow$. 
This is different from the DWs induced by $H_{\rm DW}$
where four types of particle-hope combination, $\langle a^\dagger_{\pm k,\chi} b^\dagger_{\pm \kb,-\chi}\rangle$,  can be non zero
simultaneously and they can give coherent effects to the correlation energy. 
Considering that the coupling constants, $G^{({ph})}$, $G^{({pp})}$, and $G^{({hh})}$ are the same order, 
the DW may be superior to the BCS-type pairing in $Z=N$ systems in the present simplified 
model with a limited number of active orbits.
We consider $H_{\rm DW}$ to be the dominant term, and adopt the ansatz of $H_2=H_{\rm DW}$ and discuss the
edge DWs in the Hamiltonian $H=H_0+H_1+H_{\rm DW}$ in relation to the SSB.

\section{Alternative method to solve DW Hamiltonian}\label{sec:meanField}
In this section, we solve Eq.~(\ref{eq:DWHamiltonian}) in the approximation omitting the quantum fluctuation of the product of the 
particle and hole operators. It is a kind of the mean-field approaches in the field theory. 
We show that the same result as that in Sec.~\ref{sec:discussion} are obtained in this approximation.

Let us consider the $H=H_0+H_1+H_\text{DW}$ in Eq.~(\ref{eq:DWHamiltonian}). By decomposing $H_\text{DW}$
into mean fields and their fluctuations, we can rewrite the second and third terms  
 $H_1+H_\text{DW}$ as
\begin{align}
H_1+H_{\rm DW}&= H_\text{mf}+H_\text{quasi}+H_\text{fluc},\notag\\
H_\text{mf}&\equiv-2\sum_{\chi , \chi'}G^{({ph})}_{\chi,\chi'}\Bigl[
\langle a^\dagger_{+k,\chi} {b}^\dagger_{+\kb,-\chi}\rangle\langle {b}_{+\kb,-\chi'} a_{+k,\chi'}\rangle+\langle a^\dagger_{-k,\chi} {b}^\dagger_{-\kb,-\chi}\rangle \langle {b}_{-\kb,-\chi'} a_{-k,\chi'}\rangle
\Bigr],\notag\\
H_\text{quasi}&\equiv H_1+
2\sum_{\chi , \chi'}G^{({ph})}_{\chi,\chi'}\Bigl[
a^\dagger_{+k,\chi} {b}^\dagger_{+\kb,-\chi}\langle {b}_{+\kb,-\chi'} a_{+k,\chi'}\rangle+\langle a^\dagger_{+k,\chi} {b}^\dagger_{+\kb,-\chi}\rangle{b}_{+\kb,-\chi'}a_{+k,\chi'} \notag\\
&\quad\qquad\qquad\quad+a^\dagger_{-k,\chi} {b}^\dagger_{-\kb,-\chi}\langle {b}_{-\kb,-\chi'} a_{-k,\chi'}\rangle+\langle a^\dagger_{-k,\chi} {b}^\dagger_{-\kb,-\chi}\rangle{b}_{-\kb,-\chi'}a_{-k,\chi'} 
\Bigr], \notag\\
H_\text{fluc}&\equiv2\sum_{\chi , \chi'}G^{({ph})}_{\chi,\chi'}\Bigl[
\bigl(a^\dagger_{+k,\chi} {b}^\dagger_{+\kb,-\chi} -\langle a^\dagger_{+k,\chi} {b}^\dagger_{+\kb,-\chi}\rangle\bigr)\bigl( {b}_{+\kb,-\chi'} a_{+k,\chi'}-\langle {b}_{+\kb,-\chi'} a_{+k,\chi'}\rangle\bigr)\notag\\
&\quad\quad\qquad\qquad+\bigl(a^\dagger_{-k,\chi} {b}^\dagger_{-\kb,-\chi} -\langle a^\dagger_{-k,\chi} {b}^\dagger_{-\kb,-\chi} \rangle\bigr)\bigl( {b}_{-\kb,-\chi'} a_{-k,\chi'}-\langle {b}_{-\kb,-\chi'} a_{-k,\chi'}\rangle\bigl)\Bigr],
\end{align}
where $H_\text{mf}$,  $H_\text{quasi}$, and $H_\text{int}$  are the mean-field energy term, the sum of $H_1$ and
the interaction term between particles and the mean field, and the fluctuation term of the mean field, respectively.
In the mean-field approximation, $H_\text{fluc}$ is assumed to be negligible, so we drop the $H_\text{fluc}$.
We also assume that the ground state does not break time-reversal symmetry, so that the mean fields satisfy
\begin{equation}\label{eq:timereversal}
\langle {b}_{+\kb,-\chi}a_{+k,\chi}\rangle=
-\langle a_{-k,-\chi}^\dag{b}_{-\kb,\chi}^\dag\rangle .
\end{equation}
The Hamiltonian of the quasiparticle $H_\text{quasi}$ can be written as
\begin{align}\label{eq:Hquasi}
H_\text{quasi}&=
\sum_\chi\begin{pmatrix}
a_{+k,\chi}^\dag & {b}_{+\kb,\chi} & a_{-k,\chi}^\dag&{b}_{-\kb,\chi} 
\end{pmatrix}
\begin{pmatrix}
E_{k,\tau} & \Delta_{\chi} & 0 & 0\\
\Delta^*_{\chi} &E_{\kb,\tau} & 0 & 0\\
0 & 0 & E_{k,\tau}&-\Delta_{-\chi}^*\\ 
0 & 0 & -\Delta_{-\chi} &E_{\kb,\tau}
\end{pmatrix}
\begin{pmatrix}
a_{+k,\chi}\\
{b}_{+\kb,\chi}^\dag\\
a_{-k,\chi}\\
{b}_{-\kb,\chi}^\dag
\end{pmatrix}
-2\sum_{\chi} E_{\kb,\tau},
\end{align}
where the last term in Eq.~(\ref{eq:Hquasi}) comes from the anticommutation relation: ${b}_{\pm \kb,\chi}^\dag {b}_{\pm \kb,\chi}=-{b}_{\pm \kb,\chi}{b}^\dag_{\pm \kb,\chi}+1$.
The gap is defined by
\begin{equation}\label{eq:consistencyCondition}
\Delta_{\chi}\equiv2\sum_{\chi'}G_{\chi,\chi'}\langle {b}_{\kb,-\chi'} a_{ k,\chi'} \rangle.
\end{equation}
Since $H_\text{quasi}$ depends on $\Delta_{\chi}$, the right handed side in Eq.~(\ref{eq:consistencyCondition}) also depends on $\Delta_{\chi}$ through the expectation value; thus, Eq.~(\ref{eq:consistencyCondition}) can be regarded as a self-consistency equation.
The Hamiltonian of the quasiparticle has the quadratic form,
 so that it can be diagonalized by the following unitary transformation or Bogoliubov transformation:
\begin{align}
&\quad
\begin{pmatrix}
\tilde a_{+ k,\chi} \\
\tilde {b}^\dag_{+\kb,\chi}
\end{pmatrix}=
\begin{pmatrix}
v_{\chi} & -u_{\chi}\\
u^*_{\chi} & v^*_{\chi}
\end{pmatrix}
\begin{pmatrix}
a_{+k,\chi} \\
{b}^\dag_{+k,\chi}
\end{pmatrix}, 
\notag\\
&\quad
\begin{pmatrix}
\tilde a_{- k,\chi} \\
\tilde {b}^\dag_{-\kb,\chi}
\end{pmatrix}=
\begin{pmatrix}
v^*_{\chi} & u^*_{\chi}\\
-u_{\chi} & v_{\chi}
\end{pmatrix}
\begin{pmatrix}
a_{-k,\chi} \\
{b}^\dag_{-k,\chi}
\end{pmatrix},
\end{align}
with
\begin{align}\label{eq:eigenvector}
\frac{u_{\chi}}{v_{\chi} }&= \frac{-2\Delta_{\chi}}{E_{k,\tau}-E_{\kb,\tau}+\sqrt{(E_{k,\tau}-E_{\kb,\tau})^2+4|\Delta_{\chi}|^2}},
\end{align}
which satisfies $|v_{\chi}|^2+|u_{\chi}|^2=1$. 
The eigenvalues of $H_\text{quasi}$ corresponding to the energies of the quasiparticles are
\begin{align}
\tilde E_{k,\chi}(\Delta_{\chi})&=\frac{1}{2}\Bigl(E_{k,\tau}+E_{\kb,\tau}+\sqrt{(E_{k,\tau}-E_{\kb,\tau})^2+4|\Delta_{\chi}|^2}\Bigr),\notag\\
\tilde E_{\kb,\chi}(\Delta_{\chi})&=\frac{1}{2}\Bigl(E_{k,\tau}+E_{\kb,\tau}-\sqrt{(E_{k,\tau}-E_{\kb,\tau})^2+4|\Delta_{\chi}|^2}\Bigr).
\end{align}
The new vacuum is defined as the state vanished by the annihilation operators, $\tilde a_{\pm k,\chi}$ and $\tilde{b}_{\pm\kb,\chi}$:
\begin{equation}\label{eq:vacuumCondition}
\tilde a_{\pm k,\chi}|\Psi\rangle=\tilde {b}_{\pm\kb,\chi}|\Psi\rangle=0 .
\end{equation}
The solution of Eq.~(\ref{eq:vacuumCondition}) is given by
\begin{equation}\label{eq:waveFunctionMeanField}
|\Psi(\Delta_\chi)\rangle
=\prod_\chi(v_{\chi}+u_{\chi}  a^\dag_{+k,\chi}{b}^\dag_{+k,\chi})\prod_\chi(v_{\chi}^*-u_{\chi}^*  a^\dag_{-k,\chi}{b}^\dag_{-k,\chi})|0\rangle_\text{F}.
\end{equation}
This has the same form as Eq.~(\ref{eq:waveFunction}); however, they are different, because 
the vacuum in Eq.~(\ref{eq:waveFunctionMeanField}) is a function of $\Delta_\chi$, while that in Eq.~(\ref{eq:waveFunction}) is 
a function of $u_\tau$ (and $v_\tau$), which is a variational parameter. Their vacua coincide at the solutions of  the self-consistent equation and the variational equation.

The expectation value of $H_\text{quasi}$ becomes
\begin{equation}\label{eq:Hquas2}
E_\text{quasi}(\Delta_{\chi})=\langle H_\text{quasi}\rangle=2\sum_{\chi}\bigl(\tilde E_{\kb,\chi}(\Delta_\chi)-E_{\kb,\chi}\bigr),
\end{equation}
where $E_\text{quasi}(\Delta_{\chi})\leq0$ and the equality is only satisfied, when $|\Delta_\chi|=0$.
The mean-field term can be rewritten by $\Delta_{\chi}$ as
\begin{align}\label{eq:Hmf2}
H_\text{mf}&=-\sum_{\chi, \chi'}\frac{ 1-3\delta_{\chi,\chi'} }{3g^{(ph)}}\Delta^*_{\chi}\Delta_{\chi'},
\end{align}
where we used the explicit form of the interaction $G^{(ph)}_{\chi\chi'}=g^{(ph)}(1-\delta_{\chi,\chi'})$.
Using Eqs.~(\ref{eq:Hquas2}) and (\ref{eq:Hmf2}),
we obtain the correlation energy as
\begin{align}\label{eq:ecorrmf}
2E_\text{corr}(\Delta_{\chi})&= E_\text{quasi} +H_\text{mf}\notag\\
&=2\sum_\chi\Bigl(\tilde E_{\kb,\chi}(\Delta_\chi)-E_{\kb,\chi} -\sum_{\chi'}\frac{ 1-3\delta_{\chi,\chi'} }{6g^{(ph)}}\Delta_{\chi}^*\Delta_{\chi'}\Bigr).
\end{align}
In the mean-field approximation, $\Delta_\chi$ is obtained by the stationary condition:
\begin{align}\label{eq:GapEq}
\frac{\partial}{\partial\Delta^*_{\chi}}E_\text{corr}(\Delta_{\chi})&=
\frac{-\Delta_{\chi}}{\sqrt{(E_{ k,\tau}-E_{ \kb,\tau})^2+4|\Delta_{\chi}|^2}}+\frac{\Delta_{\chi}}{2g^{(ph)}}-\frac{\sum_{\chi'}\Delta_{\chi'}}{6g^{(ph)}}
=0.
\end{align}
Notice that Eq.~(\ref{eq:GapEq}) is the equivalent to the consistency condition Eq.~(\ref{eq:consistencyCondition}),
which can be check by inserting $\langle {b}_{\kb,-\chi} a_{ k,\chi} \rangle=u_\chi v_\chi$.

For $Z=N$ systems, when $E_{k,\tau}$ and $E_{q,\tau}$ are independent of the isospin, $\Delta_{\chi}$ is independent of $\chi$.
The solution is
\begin{equation}\label{eq:Delta}
\Delta_{\chi}=3g\sqrt{1-\frac{(E_{ k}-E_{ \kb})^2}{(6g)^2}} .
\end{equation}
Inserting Eq.~(\ref{eq:Delta}) into Eqs.~(\ref{eq:eigenvector}) and (\ref{eq:ecorrmf}), one finds that the $v_\tau$ and $u_\tau$ 
coincide with Eq.~(\ref{eq:theSolutionOfMeanField}) and also the $E_{\rm corr}$ does with Eq.~(\ref{eq:ecorrvp}). 

\section*{Acknowledgments}
The computational calculations of this work were performed by using the
supercomputers at YITP and done in Supercomputer Projects 
of High Energy Accelerator Research Organization (KEK).
This work was supported by Grant-in-Aid for Scientific Research from Japan Society for the Promotion of Science (JSPS).
It was also supported by 
the Grant-in-Aid for the Global COE Program ``The Next Generation of Physics, 
Spun from Universality and Emergence'' from the Ministry of Education, Culture, Sports, Science and Technology (MEXT) of Japan. 
Discussions during the YITP workshops and the YIPQS long-term workshops held at YITP are 
helpful to complete this work.


\section*{References}
\begin{thebibliography}{9}
\bibitem{brink66} D. M. Brink, International School of Physics ``Enrico Fermi'', XXXVI, p. 247 (1966).

\bibitem{yukawa70} T. Yukawa and S. Yoshida, Phys. Lett. B 33, 334 (1970).

\bibitem{horiuchi87} H. Horiuchi and K. Ikeda, in Cluster Models and Other Topics, International Review of Nuclear Physics (World Scientific, Singapore, 1987), Vol. 4, p. 1. 

\bibitem{nucldata}
F. Ajzenberg-Selove, Nucl. Phys. A {\bf 506}, 1 (1990).

\bibitem{uegaki77} E. Uegaki, S. Okabe, Y. Abe, and H. Tanaka, Prog. Theor. Phys. {\bf 54}, 1262 (1977).

\bibitem{glatz81} F. Glatz, et al., Z. Physics. {\bf A 303}, 239 (1981). 

\bibitem{bauhoff82} W. Bauhoff, Z. Phys. {\bf A305}, 187 (1982). 

\bibitem{bauhoff82b} W. Bauhoff, et al., Phys. Ref. C {\bf 26},1725 (1982). 

\bibitem{Maruhn:2006ig}
  J.~A.~Maruhn, M.~Kimura, S.~Schramm, P.~G.~Reinhard, H.~Horiuchi and A.~Tohsaki,
  Phys.\ Rev.\  C {\bf 74}, 044311 (2006).

\bibitem{KanadaEn'yo:2004cv}
  Y.~Kanada-En'yo,
  Phys.\ Rev.\  C {\bf 71}, 014303 (2005).

\bibitem{KanadaEn'yo:2003di}
  Y.~Kanada-En'yo, M.~Kimura and H.~Horiuchi,
  Nucl.\ Phys.\  A {\bf 734}, 341 (2004).
\bibitem{overhauser60}
A. W. Overhauser, Phys. Rev. Lett. {\bf 4}, 415 (1960).
\bibitem{llano79} M. de Llano, Nucl. Phys. A {\bf 317}, 183 (1979).
\bibitem{ui81} H. Ui and Y. Kawazore, Z. Phys. A {\bf 301}, 125 (1981). 
\bibitem{brink73}  D. M. Brink and J. J. Castro, Nucl. Phys. {\bf A216}, 109 (1973).
\bibitem{tohsaki89} A. Tohsaki-Suzuki, Prog. Theor. Phys. {\bf 81}, 370 (1989).
\bibitem{takemoto04} H. Takemoto, Phys. Rev. C {\bf 69}, 035802 (2004).

  
  
\bibitem{ENYOabc}
  Y.~Kanada-Enyo and H.~Horiuchi,
  Prog.\ Theor.\ Phys.\  {\bf 93}, 115 (1995);
  Y.~Kanada-Enyo, H.~Horiuchi and A.~Ono,
  Phys.\ Rev.\  C {\bf 52}, 628 (1995);
  Y.~Kanada-Enyo and H.~Horiuchi,
  Phys.\ Rev.\  C {\bf 52}, 647 (1995).
  
\bibitem{ENYOsupp}
  Y. Kanada-En'yo and H. Horiuchi,
  Prog. Theor. Phys. Suppl. {\bf 142}, 205 (2001).
\bibitem{AMDrev} 
  Y. Kanada-En'yo M. Kimura and H. Horiuchi, 
  C. R. Physique {\bf 4} 497 (2003).

\bibitem{TOHSAKI}
 T. Ando, K.Ikeda and A. Tohsaki, Prog. Theor. Phys.
 {\bf 64}, 1608 (1980).
\bibitem{LS}
 N. Yamaguchi, T. Kasahara, S. Nagata and Y. Akaishi,
 Prog. Theor. Phys. {\bf 62}, 1018 (1979);
 R. Tamagaki, Prog. Theor. Phys. {\bf 39}, 91 (1968).





\bibitem{DOTE-be}
A. Dot\'e, H. Horiuchi and Y. Kanada-En'yo,
Phys. Rev. C {\bf 56}, 1844 (1997).

\bibitem{BCS}
J.~Bardeen, L.~N.~Cooper and J.~R.~Schrieffer, Phys. Rev. {\bf 108}, 1175 (1957).  


\bibitem{tamagaki76}
R. Tamagaki and T. Takatsuka, Prog. Theor. Phys. {\bf 56},1340 (1976).
\bibitem{takatsuka78}
T. Takatsuka, K. Tamiya, T. Tatsumi and R. Tamagaki, Prog. Theor. Phys. {\bf 59}, 1933 (1978). 
\bibitem{migdal78}
A. B. Migdal, Rev. Mod. Phys. {\bf 50}, 107 (1978).

\bibitem{Dautry:1979bk}
  F.~Dautry and E.~M.~Nyman,
  Nucl.\ Phys.\  A {\bf 319}, 323 (1979).

\bibitem{Deryagin:1992rw}
  D.~V.~Deryagin, D.~Y.~Grigoriev and V.~A.~Rubakov,
  Int.\ J.\ Mod.\ Phys.\  A {\bf 7}, 659 (1992).

\bibitem{Shuster:1999tn}
  E.~Shuster and D.~T.~Son,
  Nucl.\ Phys.\  B {\bf 573}, 434 (2000).

\bibitem{Park:1999bz}
  B.~Y.~Park, M.~Rho, A.~Wirzba and I.~Zahed,
  Phys.\ Rev.\  D {\bf 62}, 034015 (2000).

\bibitem{Alford:2000ze}
  M.~G.~Alford, J.~A.~Bowers and K.~Rajagopal,
  Phys.\ Rev.\  D {\bf 63}, 074016 (2001).


\bibitem{Nakano:2004cd}
  E.~Nakano and T.~Tatsumi,
  Phys.\ Rev.\  D {\bf 71}, 114006 (2005).
\bibitem{Giannakis:2004pf}
  I.~Giannakis and H.~C.~Ren,
  Phys.\ Lett.\  B {\bf 611}, 137 (2005).

\bibitem{Fukushima:2006su}
  K.~Fukushima,
  Phys.\ Rev.\  D {\bf 73}, 094016 (2006).

\bibitem{Nickel:2009ke}
  D.~Nickel,
  Phys.\ Rev.\ Lett.\  {\bf 103}, 072301 (2009);
  Phys.\ Rev.\  D {\bf 80}, 074025 (2009).
 
\bibitem{Kojo:2009ha}
  T.~Kojo, Y.~Hidaka, L.~McLerran and R.~D.~Pisarski,
  Nucl.\ Phys.\  A {\bf 843}, 37 (2010).

\bibitem{Carignano:2010ac}
  S.~Carignano, D.~Nickel and M.~Buballa,
  Phys.\ Rev.\  D {\bf 82}, 054009 (2010).



  

\bibitem{Fukushima:2010bq}
  K.~Fukushima, T.~Hatsuda,
  Rept.\ Prog.\ Phys.\  {\bf 74}, 014001 (2011).


\bibitem{CDW}
 G. Gruner, Rev. Mod. Phys. 60, 1129 (1988). 
 \bibitem{SDW}
G. Gruner, Rev. Mod. Phys. 66, 1 (1994).



\end{thebibliography}
\end{document}